\documentclass[aps,prb,twocolumn,groupedaddress,superscriptaddress,nofootinbib,amsmath,amssymb,floats,floatfix,english]{revtex4-2}

\usepackage[utf8]{inputenc}
\usepackage[T1]{fontenc}
\usepackage{amsmath}
\usepackage{amssymb}
\usepackage{ifthen}
\usepackage{graphicx}
\usepackage{times}
\usepackage{babel}
\usepackage{color}
\usepackage{bm}
\usepackage[normalem]{ulem}
\usepackage{float}
\usepackage{placeins}
\usepackage{chngcntr}
\usepackage{enumitem}
\usepackage{multirow}
\usepackage{mathtools}

\usepackage{comment}

\newcommand{\beq}{\begin{equation}}
\newcommand{\eeq}{\end{equation}}
\newcommand{\beqy}{\begin{eqnarray}}
\newcommand{\eeqy}{\end{eqnarray}}
\newcommand{\beqs}{\begin{equation*}}
\newcommand{\eeqs}{\end{equation*}}
\newcommand{\bpm}{\begin{pmatrix}}
\newcommand{\epm}{\end{pmatrix}}

\begin{document}

\title{Antibunching in locally driven dissipative Lieb lattices}

\author{Alex Ferrier}
\email{a.ferrier@ucl.ac.uk}
\affiliation{Department of Physics and Astronomy, University College London, Gower Street, London WC1E 6BT, United Kingdom}
\affiliation{Center for Theoretical Physics, Polish Academy of Sciences, Aleja Lotnik\'ow 32/46, 02-668 Warsaw, Poland}

\author{Michał Matuszewski}
\affiliation{Center for Theoretical Physics, Polish Academy of Sciences, Aleja Lotnik\'ow 32/46, 02-668 Warsaw, Poland}

\author{Piotr Deuar}
\affiliation{Institute of Physics, Polish Academy of Sciences, Aleja Lotnik\'ow 32/46, 02-668 Warsaw, Poland}

\author{Marzena H. Szyma\'nska}
\affiliation{Department of Physics and Astronomy, University College London, Gower Street, London WC1E 6BT, United Kingdom}


\begin{abstract}
 In Lieb lattices, geometric frustration and destructive interference of hopping cancels the occupation of certain sites, leading to flat-band physics.  Here, we show numerically how, in the driven-dissipative Bose-Hubbard (DDBH) model arranged into Lieb lattices and related geometries, specific localised driving schemes can repurpose this interference to generate enhanced antibunching via a mechanism similar to the so-called unconventional photon blockade.  Stochastic simulations using the positive-P method allow us to calculate occupations and second order correlations exactly for extended lattices.  We use this to optimise the parameters for the possible observation of this effect in polariton micropillar experiments.  This work demonstrates the possibility of using localised driving and interference effects to generate non-trivial quantum correlations in open quantum lattice systems.  Specifically, producing antibunching in the dark sites of the flat band system rather than the usual and less useful bunching.   
\end{abstract}

\maketitle

\section{Introduction}

Great progress on implementing and controlling photonic systems in experiments \cite{RevModPhys.85.299} has led to a wide array of platforms for realising open quantum systems.  Cavity QED \cite{Raimond01r,Walther06r,Reiserer15r}, circuit QED \cite{Schmidt13,Houck12,Fink17,Fitzpatrick17,Kollar19}, arrays of optical cavities \cite{Carusotto09,Umcal12}, quantum dots \cite{Kasprzak2010}, polariton lattices \cite{Amo16r,Schneider16,Lai07,Kim11,Tanese2013,Tanese14,Zhang15,Baboux16,St-Jean17,Klembt17,Klembt2018,Whittaker18,Goblot19,Milicevic19,Su20,Dang20,Dusel20}, and ultracold atoms \cite{Brennecke2007}, have each found success as physical implementations of various driven-dissipative quantum models.  While significant external drive and dissipation can normally be considered a detriment to maintaining non-trivial quantum correlations, it also endows these systems with a high degree of accessibility and tunability of their internal states.  Quantum features of stationary states of an open system do not suffer from the fragility inherent to closed systems.  Likely due to the vast number of possible configurations that can be considered, the possibilities enabled by engineering specifically localised driving schemes in interacting open quantum systems have only recently begun to be studied \cite{Usaj24,PhysRevA.109.063523}.  

Many of these systems can also be engineered into a variety of lattice geometries, which may carry their own geometric or topological properties.  Lieb lattices have been widely studied with polariton micropillars \cite{Baboux16,Klembt17,Whittaker18,Goblot19}, which we will \mbox{focus} on as the primary physical context for the investigations performed in this work.  They can display a range of interesting physics due to frustration and interference effects, including a flat-band structure \cite{Biondi15,PhysRevA.87.023614}. It was previously shown that in the mode corresponding to the occupation of the flat band, interference in driven-dissipative Lieb lattices results in a strong bunching of second order correlations on the less occupied ``dark'' sites of each unit cell \cite{Casteels16,PRXQuantum.2.010319}, which makes a large departure from coherent or thermal states.  

In contrast to bunching, antibunching of second order correlations cannot be achieved by classical mechanisms \cite{QuantumNoise}, and so have been highly sought after as signatures of quantum correlations in exciton-polaritons \cite{Mu_oz_Matutano_2019,Delteil_2019}; however, the degree of antibunching achieved by the usual photon blockade in single micropillars is fairly modest due to the low ratio of interactions to dissipation.  Unlike the regular photon blockade, where multiple occupations are suppressed by the energy shift from interactions, the unconventional photon blockade (UPB), first discovered in \cite{Liew10}, is a mechanism that results in strong antibunching on the one driven site of a pair of coupled resonators, such as a DDBH dimer, even with these weaker interactions.  The effect is caused by destructive interference between different paths that eliminates the probability of multiple occupation of the target site; for example, in the DDBH dimer the amplitude for a second boson to occupy the coherently driven site is cancelled, either partially or completely, by the amplitude for it to hop to the neighbouring site and back \cite{Bamba11}.  Since the initial proposal, there have been a number of further investigations into different implementations of the UPB and the physics behind it \cite{Bamba11,Bamba11apl,PhysRevA.82.013841,PhysRevA.88.033836,Ferretti_2013,Liew_2013,PhysRevA.89.031803,PhysRevA.90.033809,PhysRevA.90.063824,PhysRevA.94.013815,PhysRevA.96.053810} and even some successful experimental realisations of it \cite{PhysRevLett.121.043601,PhysRevLett.121.043602}, all of which are focused on two site systems, as in the original proposal.  

In this work, we will show how on extended open quantum lattice systems the interplay plus possible underlying physical connections between the UBP and geometric frustration in Lieb lattices, along with tailored localised driving, can enable the generation of non-trivial quantum correlations on chosen sites. 
Particularly, this approach may provide an alternative avenue forward for generating significant antibunching and other quantum correlations in exciton-polariton micropillar lattices, despite their lower interaction to dissipation ratios.

\section{Model and Methods}

We investigate driven-dissipative Bose-Hubbard (DDBH) models in Lieb lattice geometries. A general form of the Hamiltonian for the system, written in the frame rotating with the driving frequency, is given by
\beqy
\hat{H} = &\sum_j \left( -\Delta\hat{a}^\dagger_j\hat{a}_j +\dfrac{U}{2}\hat{a}^\dagger_j\hat{a}^\dagger_j\hat{a}_j\hat{a}_j +F_j\hat{a}^\dagger_j + F_j^*\hat{a}_j \right) \nonumber \\ &-\sum_{j,j'} \left(J_{j,j'}\hat{a}^\dagger_j\hat{a}_{j'} + J^*_{j,j'}\hat{a}^\dagger_{j'}\hat{a}_j \right) \, , \label{hamBH}
\eeqy
where $\Delta$ and $U$ are the on-site energy detuning and interaction strength respectively, $F_j$ is the local drive on site $j$, and the hopping $J_{j,j'} = J$ for all pairs of adjacent sites, unless otherwise stated, and 0 otherwise.  Each site also experiences a local dissipation $\gamma$, such that the evolution of the density matrix $\hat{\rho}$ of the system is given by the master equation:
\beq
\frac{\partial\hat{\rho}}{\partial t} = -i\left[\hat{H}, \hat{\rho}\right] + \sum_j\frac{\gamma}{2}\left(2\hat{a}_j\hat{\rho}\hat{a}^\dagger_j - \hat{a}^\dagger_j\hat{a}_j\hat{\rho} - \hat{\rho}\hat{a}^\dagger_j\hat{a}_j\right) \, . \label{ME_DDBH} 
\eeq

To obtain the results of this work, we solve the master equation using the positive-P method.  A mapping from $\hat{\rho}$ to the probability distribution $P(\vec{\alpha})$ over a complex phase space $\vec{\alpha}\equiv \{\alpha_j,\beta_j\}$ defined by 
\beqy
\hat{\rho} = \int d\vec{\alpha} P(\vec{\alpha})\hat{\Lambda}(\vec{\alpha}) \, , \\
\hat{\Lambda}(\vec{\alpha}) = \bigotimes_j \frac{|\alpha_j\rangle\langle\beta^*_j|}{\langle\beta^*_j|\alpha_j\rangle} \, , \nonumber
\eeqy
where $\{\alpha_j,\beta_j\}$ are coherent state amplitudes defined in the usual way such that $\hat{a}_j|\alpha_j\rangle = \alpha_j |\alpha_j\rangle$, allows the master equation \eqref{ME_DDBH} to be converted to a Fokker-Planck equation for $P(\vec{\alpha})$, and in turn stochastic differential equations for trajectories of $\{\alpha_j,\beta_j\}$ which sample $P(\vec{\alpha})$.  This method allows quantum mechanical observables to be calculated from appropriate averages over the stochastic trajectories in this complex phase space.  It provides exact results in the limit of sufficiently large numbers of trajectories, provided the dissipation in the system is sufficient to stabilise the method \cite{PRXQuantum.2.010319,Naether15,Mandt15}.  Further details of how this method is implemented for the DDBH model are provided in our previous work \cite{PRXQuantum.2.010319}.  The resulting Ito stochastic differential equations for the complex phase space variables $(\alpha_j,\beta_j)$ are
\begin{subequations}\label{DDBHpp_Ito}
\begin{align}
\frac{\partial \alpha_j}{\partial t} =& \, i\Delta\alpha_j -iU\alpha_j^2\beta_j -iF_j -\frac{\gamma}{2}\alpha_j \nonumber \\& +i\sum_{j' \neq j} J_{j,j'}\alpha_{j'}  + \sqrt{-iU} \alpha_j \xi^{(\alpha)}_j \, , \\
\frac{\partial \beta_j}{\partial t} =& \,  -i\Delta\beta_j +iU\alpha_j\beta_j^2 +iF_j^* -\frac{\gamma}{2}\beta_j \nonumber \\& -i\sum_{j' \neq j} J^*_{j,j'}\beta_{j'} + \sqrt{iU} \beta_j \xi^{(\beta)}_j \,  ,
\end{align}
\end{subequations}
where $\xi^{(z)}_j$ are uncorrelated real Gaussian noises with $\langle \xi^{(z)}_j(t) \xi^{(z')}_{j'}(t') \rangle = \delta_{z,z'}\delta_{j,j'}\delta(t-t')$ and $\langle \xi^{(z)}_j(t) \rangle = 0$. To produce trajectories, these are solved numerically using the xmds2 package \cite{XMDS_CPhysComm.184.201}.  

Relevant observables calculated in this work are the local occupations 
\beq
n_j(t) = \langle\hat{a}^\dagger_j(t)\hat{a}_j(t) \rangle = \langle \alpha_j(t)\beta_j(t) \rangle_\mathrm{PP}\, ,
\eeq
and second order correlations $g^{(2)}_j(\tau)$ with time delay $\tau$ \cite{tcorr}, which are calculated from the positive-P results as
\beqy
g^{(2)}_j(\tau) &= \dfrac{\langle \hat{a}^\dagger_j(t)\hat{a}^\dagger_j(t+\tau)\hat{a}_j(t+\tau)\hat{a}_j(t) \rangle}{n_j(t) n_j(t+\tau)} \nonumber \\ &= \dfrac{\langle \alpha_j(t)\alpha_j(t+\tau)\beta_j(t+\tau)\beta_j(t) \rangle_\mathrm{PP}}{n_j(t) n_j(t+\tau)}\, ,
\eeqy
where each average $\langle ... \rangle_{PP}$ involved in calculating these observables is taken over both the set of stochastic samples as well as over the time $t$ within the steady state.  Simulations are run starting with vacuum initial conditions (all $\alpha_j(t=0)=\beta_j(t=0)=0$), until time $t$ far into the steady state.  Averages over a total of 2000 samples are used for the results presented in  this work.

A characteristic feature of the UPB is the form of the time delayed second order correlations $g^{(2)}_j(\tau)$.  In the case of the UPB, these correlations oscillate in $\tau$, with a period inversely proportional to the hopping $J$ \cite{Liew10,Bamba11} and have antibunching at $\tau\sim 0$.  In experiments, photon detectors needed for performing measurements of $g^{(2)}_j(0)$ will typically have some finite resolution in time.  To detect the antibunching from the UPB, it is therefore useful for the period of these oscillations to be as long as possible, compared to the time resolution of the detectors.  In the following section, we will explore geometries and parameter values of the DDBH to find optimal conditions for future experiments to observe the correlations.  

\section{Results}

\subsection{Optimal parameters for antibunching in a three site chain}\label{section:3siteUPB}

\begin{figure}[t]
\begin{center}
\includegraphics[width = 0.8\columnwidth]{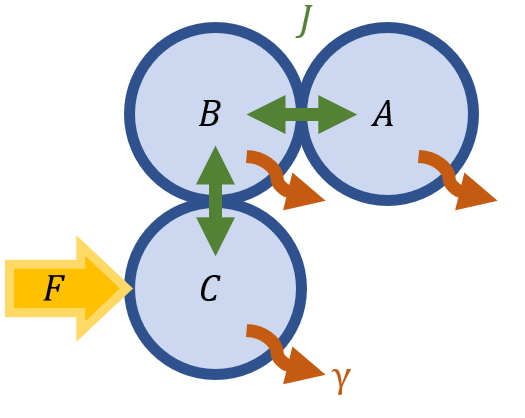}
\caption{Diagram of a single Lieb lattice unit cell, equivalent to a \mbox{3 site} 1D chain with open boundaries.  Local dissipation $\gamma$ occurs equally at all sites, but a coherent drive $F$ is applied only to the $C$ site, with hopping $J$ between neighbouring sites allowing occupation to then spread to other sites in the lattice.  \label{Lieb1}}
\end{center}
\end{figure}

We begin by considering the simplest example relevant to Lieb lattices, a single unit cell or three site 1D chain, shown in Fig.~\ref{Lieb1}.  This system has the advantage of being small enough that it is still possible to make progress with some of the analytical techniques that have been used to study the two site UPB.  In particular, we take the method of \cite{Bamba11}, where the optimal parameters for UPB are found by solving the master equation in the weak driving limit when setting the amplitude for two bosons to occupy relevant site to 0, and apply it to the three site chain, where the amplitude we set to 0 in this case is that for two bosons to occupy the middle ($B$) site.  

This process, which we detail fully in Appendix \ref{appendix:3siteUPB}, leads to a pair of equations which relate the values of physical parameters that produce optimal antibunching of the middle site,
\begin{subequations}\label{opteq}
\begin{align}
4\Delta^3 -U\Delta^2 -3\gamma^2\Delta -J^2U +\frac{U\gamma^2}{4} = 0 \, , \label{opteqRe} \\
6\gamma\Delta^2 -U\gamma\Delta - \frac{\gamma^3}{2} = 0 \, . \label{opteqIm} 
\end{align}
\end{subequations}
We identify $\Delta_{opt}$ and $J_{opt}$ as the values of $\Delta$ and $J$ that give optimal antibunching, i.e.~that solve the equations \eqref{opteq}, for a given value of $U$ and $\gamma$.  Solving the quadratic equation \eqref{opteqIm} for $\Delta$ gives
\beq
\Delta_{opt} = \frac{U \pm \sqrt{U^2 +12\gamma^2}}{12} \, , \label{deltopt}
\eeq
which can then be used with the solution of \eqref{opteqRe}, to find the corresponding $J_{opt}$ as
\beq
J_{opt} = \sqrt{\frac{\gamma^2}{4} - \Delta_{opt}^2 +\frac{4\Delta_{opt}^3}{U} - \frac{3\gamma^2\Delta_{opt}}{U} } \, , \label{Jopt}
\eeq
noting that only the negative solution of \eqref{deltopt} gives real values of $J_{opt}$.  

\begin{figure}[b]
\includegraphics[width = \columnwidth]{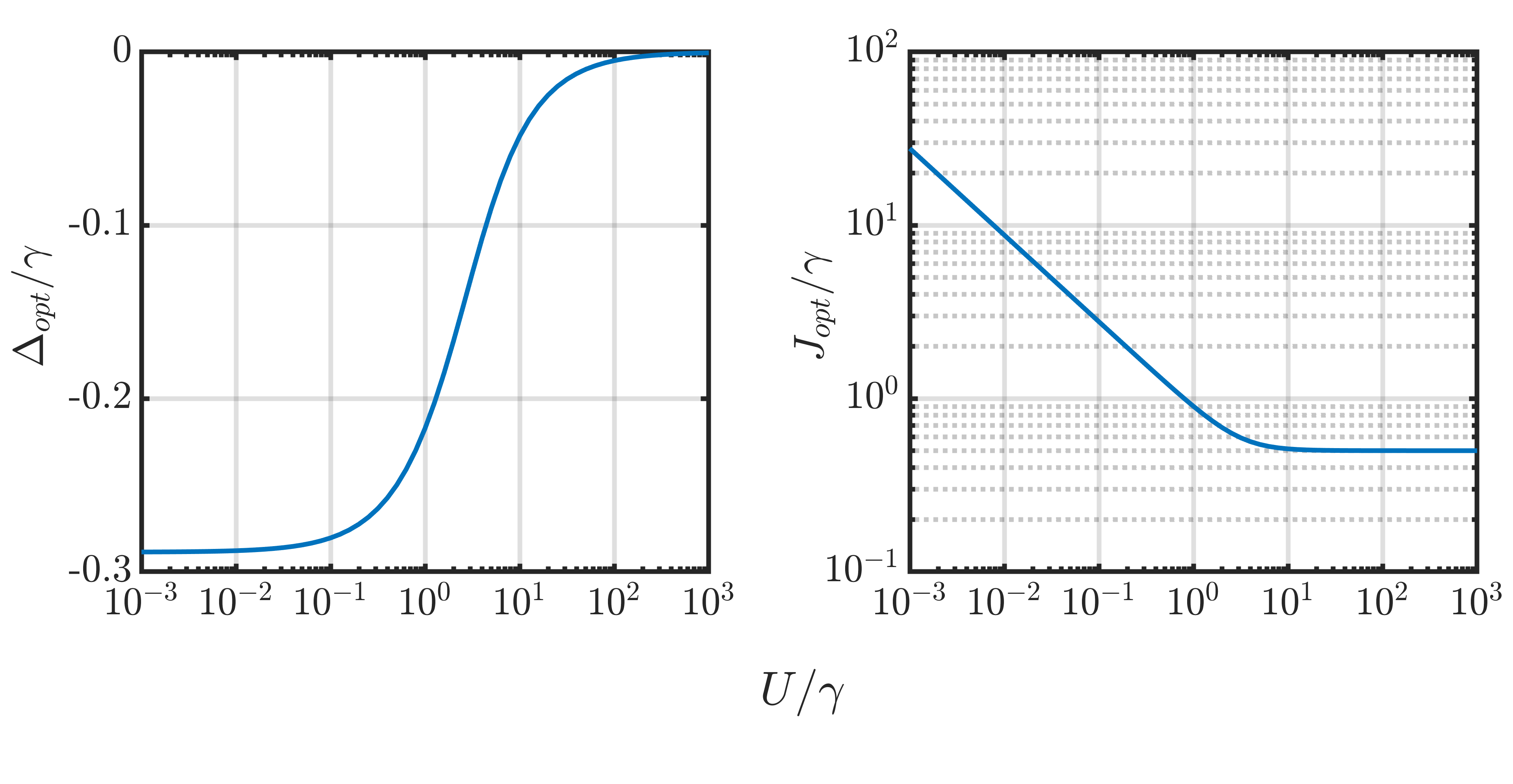}
\caption{Analytical values of optimal parameters for antibunching on central site of three site DDBH chain, calculated in the weak driving limit, across orders of magnitude in $U$.  Left panel: optimal detuning $\Delta_{opt}$.  Right panel: optimal hopping strength $J_{opt}$.  \label{3siteOpt}}
\end{figure}

The solutions of \eqref{deltopt} and \eqref{Jopt} across orders of magnitude in $U$ are shown in \mbox{Fig.~\ref{3siteOpt}}.  The behaviour is qualitatively similar to that seen for the UPB on two sites \cite{Bamba11}; however, the limit of $J_{opt}$ as $U \to \infty$ is smaller for the three site chain, $J_{opt} \to {\gamma/2}$ as opposed to $J_{opt} \to {\gamma/\sqrt{2}}$, which suggests exploring UPB-like effects in systems larger than just two sites may have advantages for increasing the period of oscillations in $g^{(2)}_j(\tau)$, which scale as $~1/J$.   

\begin{figure}[t]
\includegraphics[width = \columnwidth]{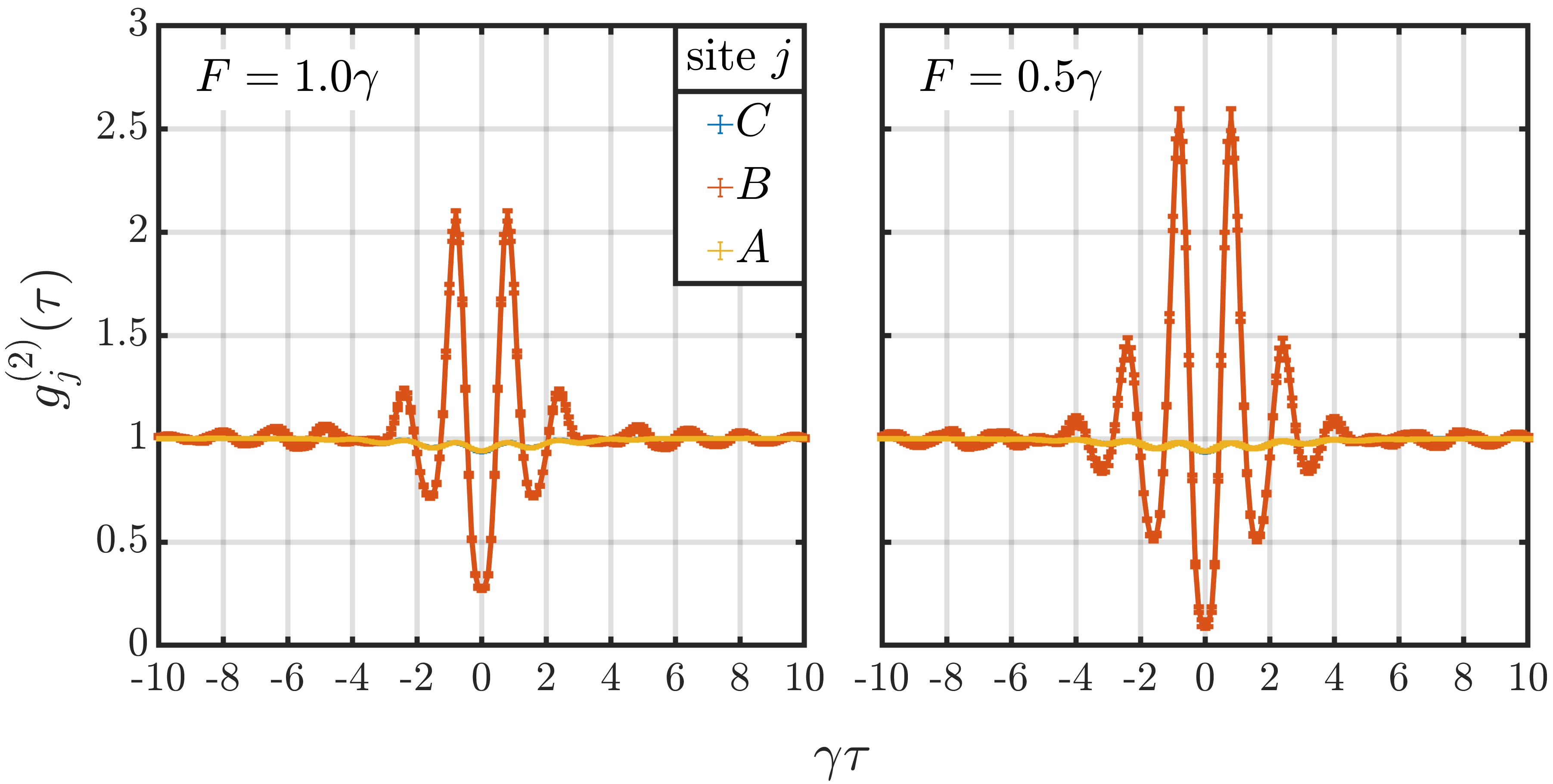}
\caption{Second order correlations $g^{(2)}_j(\tau)$ on the $B$ site using optimised parameters \mbox{$\Delta = -0.28\gamma$}, \mbox{$U = 0.1\gamma$}, \mbox{$J = 2.775\gamma$}, for two different values of the coherent drive $F$ applied to site $C$.  Note that in both cases, $g^{(2)}_j(\tau)$ for sites $A$ and $C$ are approximately equal at all $\tau$. 
\label{3sO_PPg2t}}
\end{figure}

For our positive-P simulations dedicated to this problem, we choose $U = 0.1\gamma$ as a value well within the regime of applicability of the positive-P method \cite{PRXQuantum.2.010319}, as well as accessible to the state of the art in polariton micropillar experiments.  For this value of $U$, the above formulae give $\Delta_{opt} = -0.28\gamma$ and $J_{opt} = 2.775\gamma$.  A plot of $g^{(2)}_j(\tau)$ for these optimal parameters is shown in Fig.~\ref{3sO_PPg2t}.  

The results in Fig.~\ref{3sO_PPg2t} show the characteristic oscillations with $\tau$ typically seen with the UPB.  Another feature that is apparent here is that a weaker external drive $F$ results in stronger antibunching, i.e.~smaller $g^{(2)}_2(0)$; for 
$F = \gamma$, we get \mbox{$g^{(2)}_2(0) = 0.271(6)$}, while \mbox{$F = 0.5\gamma$} results in \mbox{$g^{(2)}_2(0) = 0.10(2)$}.  There is a trade-off, however; smaller $F$ also results in reduced occupation of the relevant site, which would make the effect more difficult to observe in \mbox{experiments}.  This naturally arises as higher occupation states should be expected to have higher amplitudes for $n\geq2$ in the Fock basis, which contribute to the size of $g^{(2)}_2(0)$.

\begin{figure}[h]
\includegraphics[width = \columnwidth]{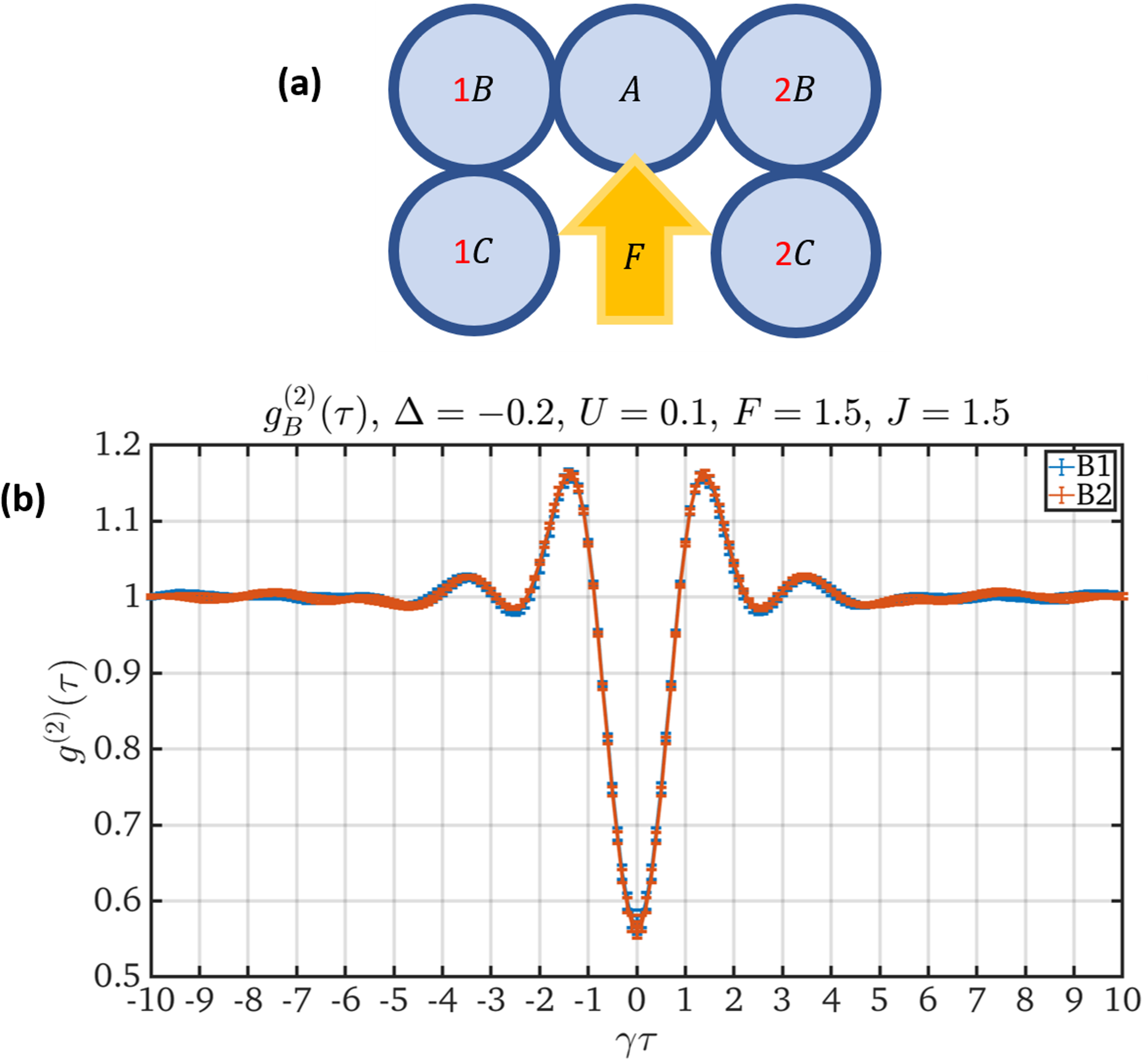}
\caption{5 site chain driven locally at central site. (a) Diagram of the structure, which we label using the conventions of the Lieb lattice. (b) Second order correlations $g^{(2)}_j(\tau)$ on the two $B$ sites for \mbox{$\Delta = -0.2\gamma$}, 
\mbox{$U = 0.1\gamma$}, \mbox{$J = F = 1.5\gamma$}; due to the symmetry, results are approximately equal for both $B$ sites.   \label{5chainAB}}
\end{figure}

\subsection{Exploring larger structures}

We now go on to explore these effects in larger structures, to investigate if any further advantage can be obtained.  The sort of analysis performed in the previous section quickly becomes unwieldy for larger lattices, so any optimisation of the physical parameters must instead be performed using the results of the positive-P simulations.  Furthermore, we direct our investigations towards maximising the viability of reproducing the results in polariton micropillar experiments; this means not just finding the smallest value of $g^{(2)}_2(0)$ achievable, but rather balancing the degree of antibunching against having large enough occupation and a large enough period of oscillations in $g^{(2)}_2(\tau)$ to make it possible to observe experimentally.  

One possible simple example beyond the single Lieb unit cell, is the 5 site chain shown in Fig.~\ref{5chainAB}.  We label the sites in this structure as if it is a quasi-1D Lieb lattice of 2 unit cells, without the $A$ site of the second unit cell.  The choice of parameters used here are not optimised, but a minimum $g^{(2)}_j(0)$ of $\approx 0.55$ is achieved on the two $B$ sites, which are adjacent to the driven site.  Compared to the results of the three site chain in Fig.~\ref{3sO_PPg2t}, the maxima of $g^{(2)}_j(\tau)$ on either side of the central minimum at $\tau = 0$ are much smaller, which could be a possible benefit to using a more complicated system such as this to achieve antibunching by this UPB-like effect.

\begin{figure}[ht]
\centering
\includegraphics[width = \columnwidth]{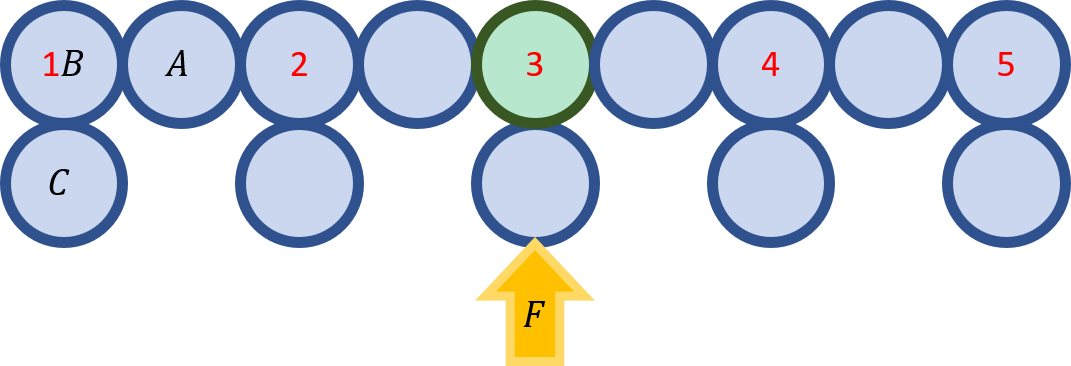}
\caption{Diagram of locally driven quasi-1D Lieb lattice of 5 unit cells with smooth edges, i.e.~there is no $A$ site on the final unit cell, with open boundary conditions.  Coherent drive is applied only to site $3C$.  The site $3B$, highlighted in green, achieves strong antibunching in $g^{(2)}(0)$ due to interference effects.  \label{LiebLpDiag}}
\end{figure}

\subsection{Antibunching in locally driven quasi-1D Lieb lattices}\label{section:LiebLp}

For the remainder of this work, we will focus on how this effect manifests in more complete Lieb lattices.  In this section, we will look at how a locally driven quasi-1D DDBH Lieb lattice can be optimised for the practical observability of antibunching occurring on a particular site.  A diagram of the exact structure we consider in these positive-P simulations is given in Fig.~\ref{LiebLpDiag}.  A coherent drive is applied only to the central site $3C$ of this 5 unit cell chain.  

\begin{figure}[ht]
\includegraphics[width = \columnwidth]{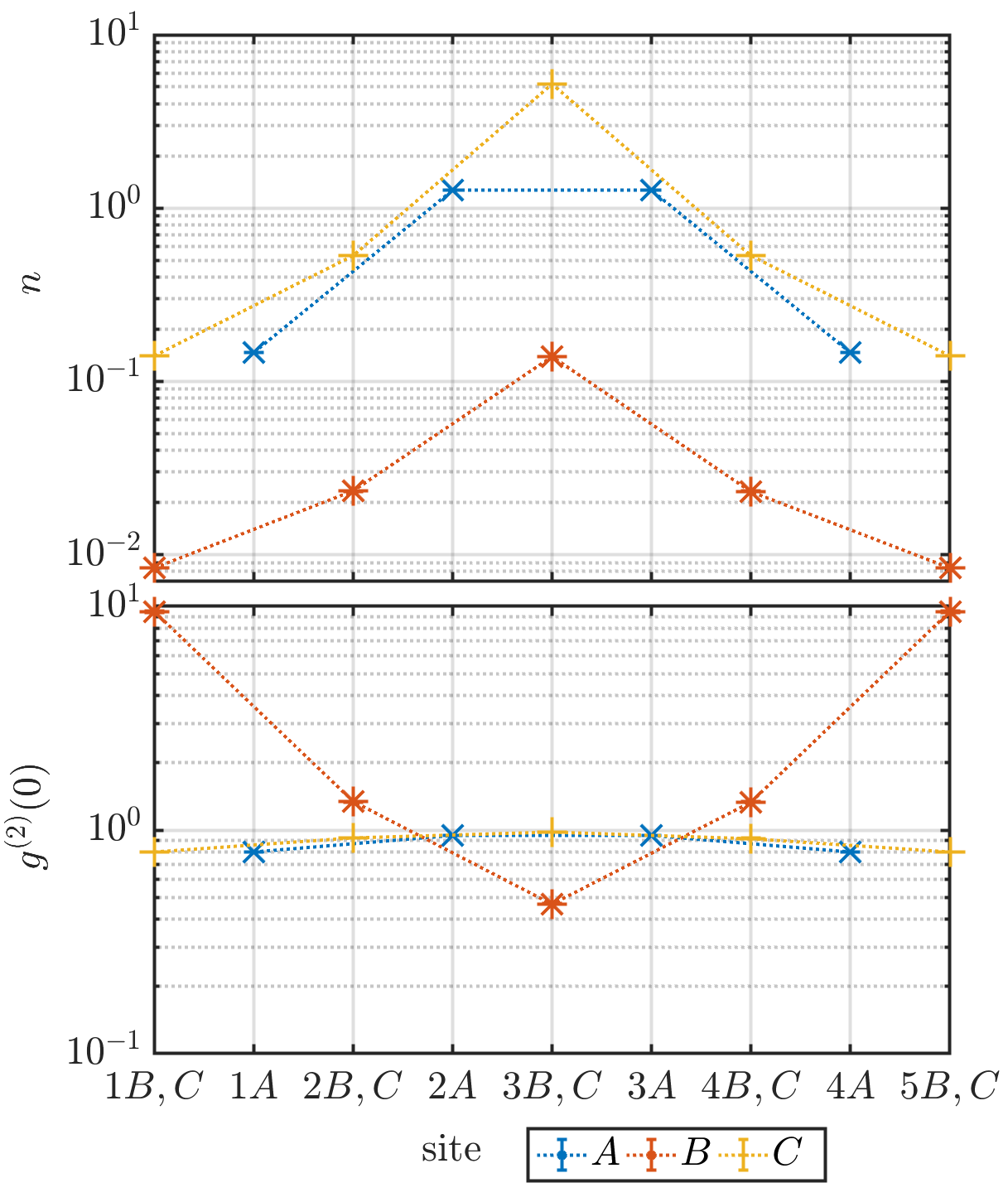}
\caption{Occupation $n$ (top panel) and second order correlation $g^{(2)}(0)$ (bottom panel) across the quasi-1D DDBH Lieb lattice of Fig.~\ref{LiebLpDiag} driven on site $3C$.  Parameters are $U = 0.1\gamma$, $\Delta = -0.2\gamma$, $F = J = 3.0\gamma$.  Values for the $A$ sites of each unit cell are horizontally offset from those for the $B$ and $C$ sites, so as to reflect the spatial structure of the lattice.  \label{LiebLpng2all}}
\end{figure}

\begin{figure}[ht]
\begin{center}
\includegraphics[width = \columnwidth]{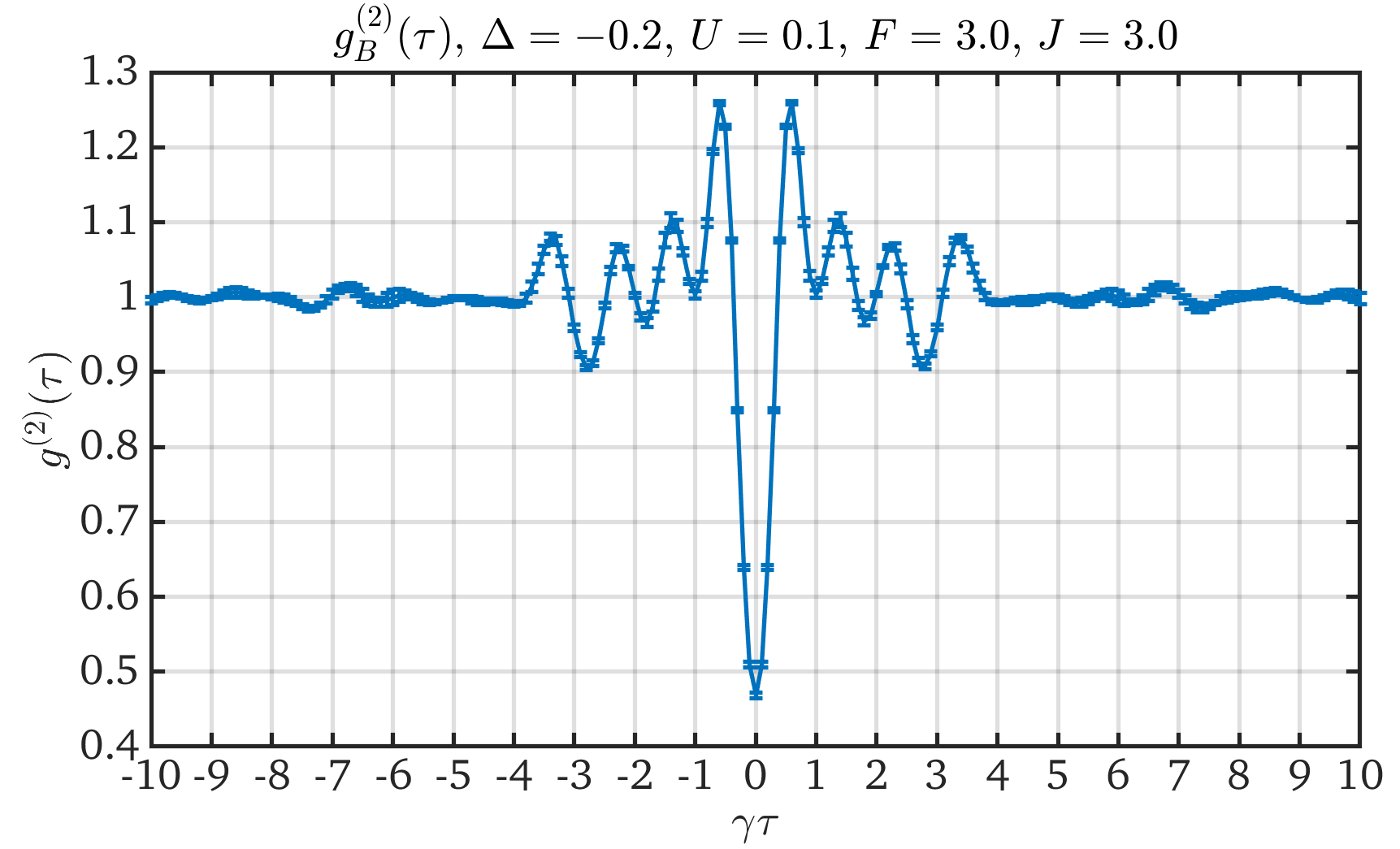}
\caption{Second order temporal correlations $g^{(2)}(\tau)$ on site $3B$ of the 5 unit cell quasi-1D DDBH Lieb lattice driven on site $3C$.  Parameters are $U = 0.1\gamma$, $\Delta = -0.2\gamma$, $F = J = 3.0\gamma$.  \label{LiebLpg2Btau}}
\end{center}
\end{figure}

Figure \ref{LiebLpng2all} shows the values of $n$ and $g^{(2)}(0)$ across the lattice that result for one example set of parameters, \mbox{$U = 0.1\gamma$,} \mbox{$\Delta = -0.2\gamma$,} \mbox{$F = J = 3.0\gamma$.}  The occupation $n$ on each of the sublattices $A,B,C$ appears to decay roughly exponentially with distance from the driven site.  Much like what is typically seen for the flat band mode of Lieb lattices \cite{Baboux16,Klembt17,Whittaker18,Goblot19,Casteels16,PRXQuantum.2.010319}, the $B$ sites are ``dark", with occupations typically around an order of magnitude less than their nearest $C$ sites.  The second order correlations show strong antibunching with \mbox{$g^{(2)}(0) = 0.468$} on the site $3B$, which is immediately adjacent the driven site, while $B$ sites further out display bunching, more like what was seen in previous work where all the $C$ sites are driven \cite{Casteels16,PRXQuantum.2.010319}.  In comparison, $A$ and $C$ sites merely show a very slight antibunching, ranging from almost completely coherent $g^{(2)}(0) \approx 1$ on the driven site to $g^{(2)}(0) \approx 0.8$ on the less occupied sites near the edges.  

As such, in what follows we focus on the values of $n$ and $g^{(2)}$ in the site $3B$ adjacent to the driven site $3C$.  Details of our investigations into optimising the system parameters for observing the antibunching on this site are given in Appendix \ref{appendix:LiebOpt}.  From that analysis we choose the parameters $\Delta = -0.2\gamma$, $U = 0.1\gamma$, $F = J = 3.0\gamma$, as a primary example that provides a good compromise between minimising $g^{(2)}(0)$ on site $3B$, while attempting to maintain enough occupation and a large enough period of $g^{(2)}(\tau)$ to make its observation realistic.  The full form of $g^{(2)}(\tau)$ on site $3B$ with these parameters is shown in Fig.~\ref{LiebLpg2Btau}.  The characteristic oscillations associated with the UPB can once again be seen, although the exact form is significantly more complicated due to the large number of sites involved.  A possible advantage to this setup is that the maxima of $g^{(2)}(\tau)$ are much smaller than in the traditional two site UPB \cite{PRXQuantum.2.010319,Bamba11,Liew10}, or the single unit cell example in Fig.~\ref{3sO_PPg2t}.  However, the period of the oscillations, which we measure as the gap in $\tau$ between the two maxima either side of $\tau = 0$, is still only about as long as the polariton lifetime ${1/\gamma}$.  A useful direction of further investigation is therefore to see if there is a way to increase this period without compromising the degree of antibunching in $g^{(2)}(0)$ achieved.

\subsection{Engineering interference with additional driving}\label{section:LiebBG}

\begin{figure}[b]
\centering
\includegraphics[width = \columnwidth]{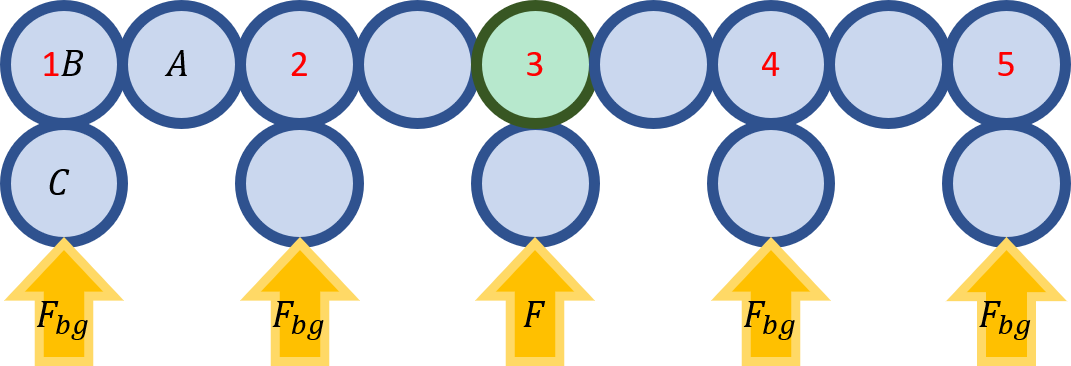}
\caption{Diagram of locally driven quasi-1D lieb lattice of 5 unit cells with smooth edges, i.e.~there is no $A$ site on the final unit cell, with open boundary conditions.  Coherent drive with strength $F$ is applied only to site $3C$, while other $C$ sites receive a coherent drive with strength $F_{bg}$.  We once again focus on the resulting behaviour of site $3B$, highlighted in green.  \label{LiebBgDiag}}
\end{figure}

In the two site UPB, applying a separate coherent drive to the second site, with an appropriate amplitude and phase \mbox{relative} to the drive on the first site, can modify the \mbox{interference} to allow optimal antibunching to be achieved for \mbox{arbitrary} values of the other parameters of the system \cite{PhysRevA.88.033836,PhysRevA.96.053810}.  Here, we investigate how a similar principle can be applied to the quasi-1D Lieb lattice.  To achieve this, we alter the scheme shown in \mbox{Fig.~\ref{LiebLpDiag}} to include an additional coherent drive, which we \mbox{refer} to as the ``background drive", at strength $F_{bg}$ on all the $C$ sites other than $3C$, which is still driven with strength $F$, as is shown in the diagram Fig.~\ref{LiebBgDiag}.   

\begin{figure}[h]
\begin{center}
\includegraphics[width = \columnwidth]{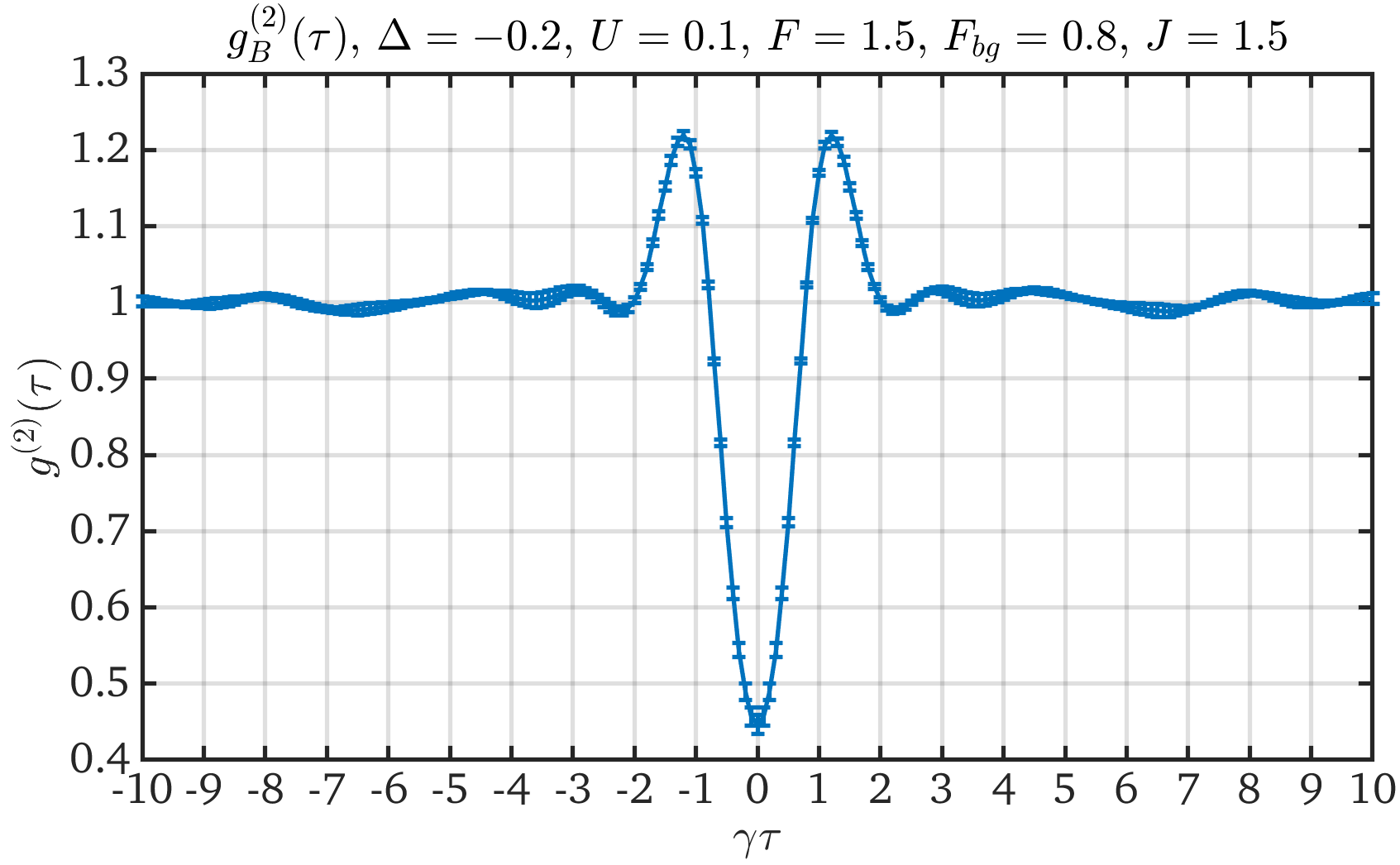}
\caption{Second order temporal correlations $g^{(2)}(\tau)$ on site $3B$ of the 5 unit cell quasi-1D DDBH Lieb lattice driven as shown in Fig.~\ref{LiebBgDiag}.  Parameters are $U = 0.1\gamma$, $\Delta = -0.2\gamma$, \mbox{$F = J = 1.5\gamma$}, \mbox{$F_{bg} = 0.8\gamma$}.  \label{LiebBgpg2Btau}} 
\end{center}
\end{figure}

The addition of this ``background drive'' $F_{bg}$ allows the \mbox{degree} of antibunching on $3B$ to be greatly improved for \mbox{parameters} that were previously suboptimal.  Specifically, in Appendix \ref{appendix:LiebOpt}, we optimise $F_{bg}$ to achieve similar results for $n$ and $g^{(2)}(0)$ at $J=1.5\gamma$.  We show in Fig.~\ref{LiebBgpg2Btau} that this decrease in $J$ leads to a proportional doubling of the period of oscillations in $g^{(2)}(\tau)$, as expected.  This has the potential to be helpful for allowing this effect to be measured in experiments, as discussed earlier.  Considering how this arises in analogy to what we observe from the analysis of the 3 site case, the additional drives should increase the amplitudes for occupation of the sites further along the lattice, offsetting the effect of the decrease in $J$ on the contribution of hopping terms from them to the target site.  In principle, even more complicated driving schemes with more independent values of the drive strength on each $C$ site or even driving on other sites could be used to further manipulate the interference effects in the lattice.  

\subsection{Relation to flat band physics}
\label{section:FB}

An interesting further direction of investigation is to \mbox{explore} the relation between this UPB-like effect and the flat band physics in Lieb lattices.  The single particle spectrum of the Lieb lattice has three bands, the middle of which has a flat dispersion \cite{Biondi15,PhysRevA.87.023614}.  The situations considered here, where the $B$ sites are dark compared to the $A$ and $C$ sites, correspond to the occupation of the flat band \cite{Casteels16,Baboux16,Klembt17,Whittaker18,Goblot19}.  

\begin{figure}[hb]
\centering
\includegraphics[width = \columnwidth]{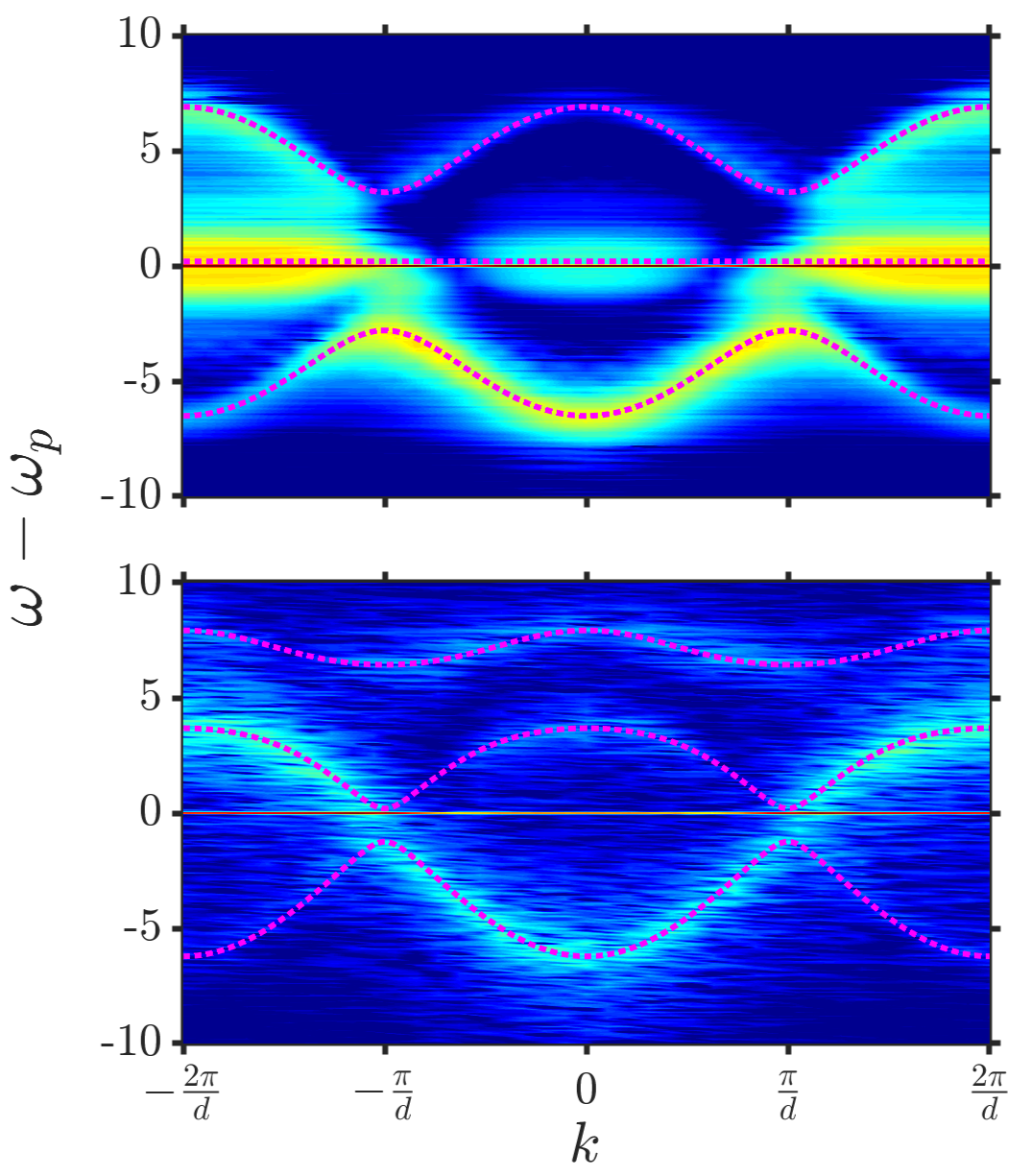}
\caption{Occupation spectrum $\tilde{n}(k, \omega)$ (colour scale, \mbox{arbitrary units}) of 20 unit cell quasi-1D Lieb lattice driven locally on site $11C$ for parameters $U = 0.1\gamma$, \mbox{$\Delta = \Delta_A = \Delta_B = -0.2\gamma$}, \mbox{$F = J = 3.0\gamma$} for 
\mbox{$\Delta_C = \Delta$} (top) and $\Delta_C = -5.0\gamma$ (bottom).  Single particle spectra in both cases are shown as dotted pink lines.  \label{specDC}}
\end{figure}

To see what role the flat band physics plays in this \mbox{enhanced} antibunching, we look at the occupation spectrum in our positive-P simulations.  This is achieved by taking the discrete Fourier transform of the stochastic complex number fields $\alpha_{S,j},\beta_{S,j}$ in space and time by
\begin{subequations}
\beqy
&\tilde{\alpha}(k, \omega) = \dfrac{1}{NN_t}\sum_{j =1}^N \sum_{t=t_0}^{t_0+(N_t-1) \delta t} e^{i\omega t} \left[ \right. \nonumber \\ &\left. \left(\alpha_{C,j} +\alpha_{B,j}\right)e^{ik(j-\frac{1}{2})d} + \alpha_{A,j}e^{ikjd}\right]\, , \\
&\tilde{\beta}(k, \omega) = \dfrac{1}{NN_t}\sum_{j =1}^N \sum_{t=t_0}^{t_0+(N_t-1) \delta t} e^{-i\omega t} \left[ \right. \nonumber \\ &\left. \left(\beta_{C,j} +\beta_{B,j}\right)e^{-ik(j-\frac{1}{2})d} + \beta_{A,j}e^{-ikjd}\right]\, , 
\eeqy
\end{subequations}
where $N$ is the number of unit cells, $d$ is the distance between equivalent lattice sites in adjacent unit cells, i.e.~twice the spacing between neighbouring sites, and $\alpha_{S,j},\beta_{S,j}$ is sampled in time at $N_t$ times evenly spaced by an interval 
$\delta t$ beginning at a time $t_0$ within the steady state.  The occupation spectrum $\tilde{n}(k, \omega)$ is then calculated by averaging over stochastic realisations
\beq
\tilde{n}(k, \omega) = \langle \tilde{\alpha}(k, \omega)\tilde{\beta}(k, \omega) \rangle_{PP} \, .
\eeq
To improve the resolution in $k$ space, we consider a modification of the setup of \mbox{Fig.~\ref{LiebLpDiag}} to include more unit cells.  \mbox{Instead} of 5 unit cells, we instead use a chain of $N = 20$ unit cells, driven on site $11C$.  Since the coherent drive is entirely \mbox{localised,} increasing the system size beyond a \mbox{certain} point should no longer noticeably affect the physics, and merely improve the $k$ resolution.  With this arrangement, we consider both the occupation spectrum $\tilde{n}(k, \omega)$ and the value of $g^{(2)}(0)$ on site $11B$.  

We probe the relation between the flat band and the \mbox{enhanced} antibunching by looking at how altering the system so that it no longer has a flat band affects the value of $g^{(2)}(0)$ on site $11B$.  One simple way to do this is to is to make the local energy detuning of the $C$ sites $\Delta_C$ different to the \mbox{local} energies of the other sites in the lattice $\Delta_A =\Delta_B = \Delta$.  We show in Fig.~\ref{specDC} the occupation spectra, overlaid with the \mbox{calculated} single particle spectra, produced in this system with $U = 0.1\gamma$, $\Delta = -0.2\gamma$, $F = J = 3.0\gamma$, for both the case with $\Delta_C = \Delta$ which exhibits a flat band, and a case with $\Delta_C = -5.0\gamma$ which transforms the flat band into a dispersive band.  For $\Delta_C = \Delta$, the site $11B$ has $g^{(2)}(0) = 0.449(9)$, while $\Delta_C = -5.0\gamma$ only gives the weak $g^{(2)}(0) = 0.948(3)$.  While the normal signature in $g^{(2)}(0)$ on the dark ($B$) sites associated with occupation of the flat band state by uniform driving of the $C$ sublattice is strong bunching $g^{(2)}(0)\gg1$ \cite{Casteels16}, by using a specifically localised drive we can repurpose the interference effects to produce antibunching $g^{(2)}(0)<1$ instead.

\subsection{Enhanced antibunching in a 2D Lieb lattice}

\begin{figure}[b]
\centering
\includegraphics[width = 0.8\columnwidth]{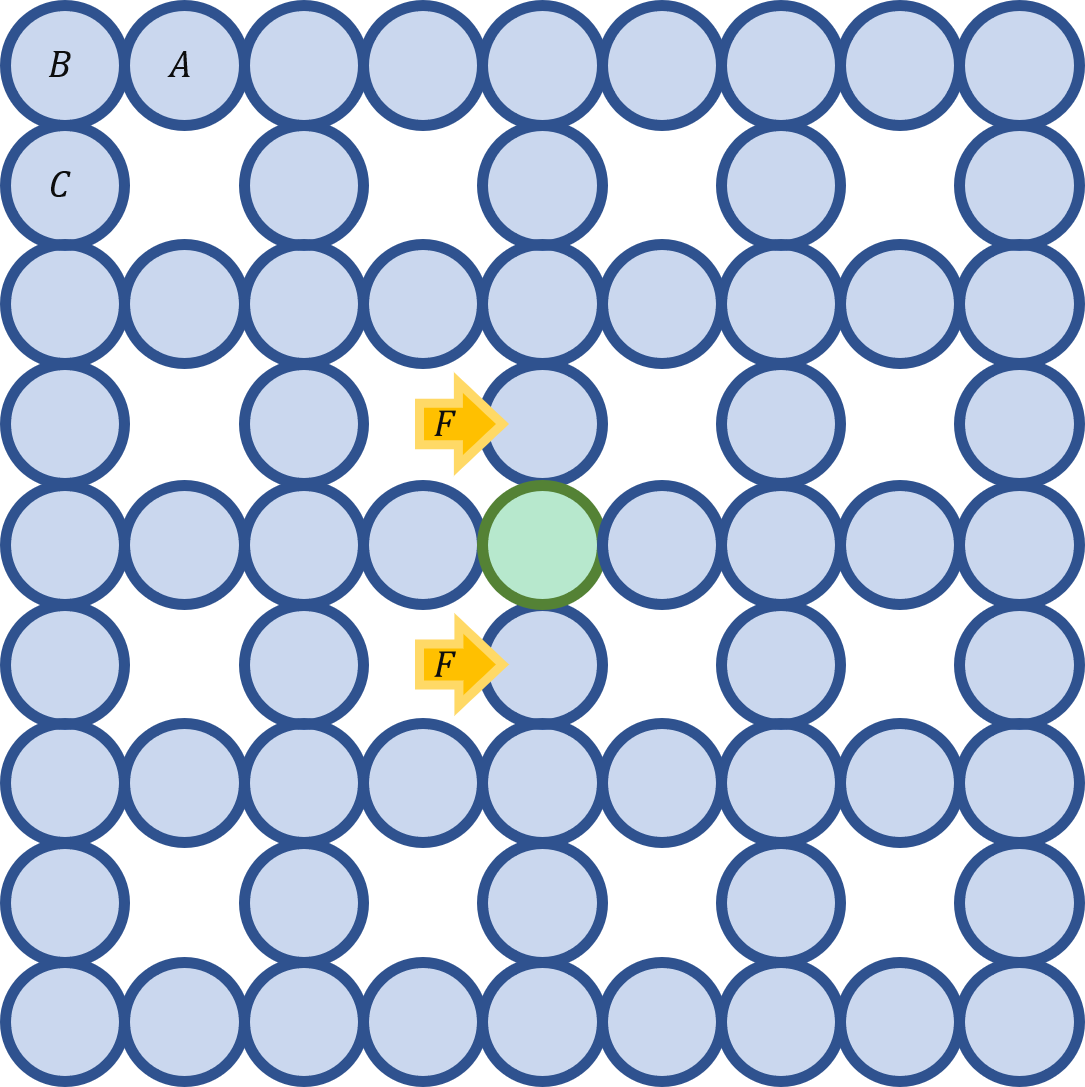}
\caption{Diagram of locally driven 2D Lieb lattice of $5\times5$ unit cells with smooth edges and open boundary conditions.  Coherent drive is applied to the two $C$ sites nearest to the centre of the lattice, $(3,3)C$ and $(3,4)C$.  The central site $(3,3)B$, highlighted in green, achieves strong antibunching in $g^{(2)}(0)$ due to interference effects.  \label{Lieb2Ddiag}}
\end{figure}

\begin{figure*}[ht]
\includegraphics[width = \textwidth]{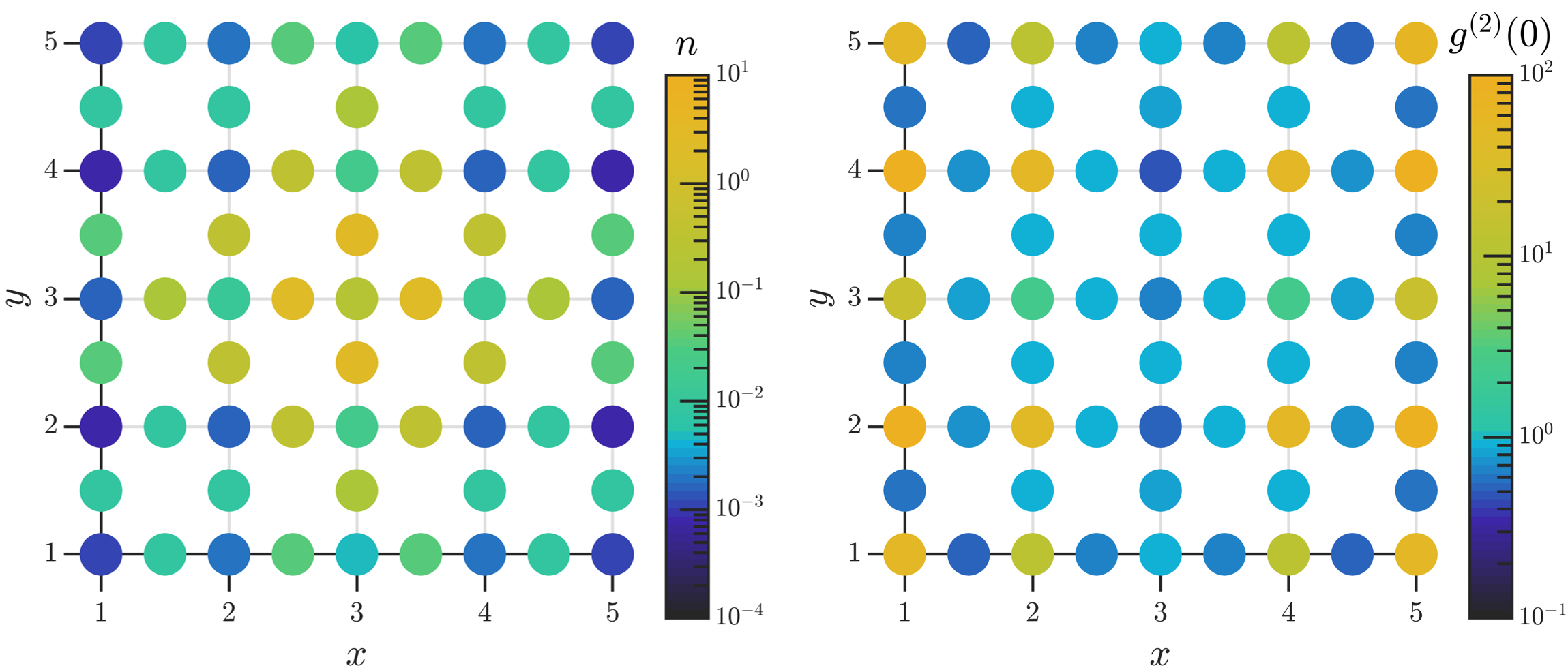}
\caption{Occupation $n$ (left panel) and second order correlation $g^{(2)}(0)$ (right panel) across a $5\times5$ unit cell 2D DDBH Lieb lattice driven on sites $(3,3)C$ and $(3,4)C$.   Parameters are $U = 0.1\gamma$, $\Delta = -0.2\gamma$, $F = J = 3.0\gamma$. \label{Lieb2Dng2}
}
\end{figure*}

Finally, we explore if this effect of enhanced \mbox{antibunching} can also be implemented in 2D Lieb lattices.  In our \mbox{exploratory} try, we found no outstanding practical advantage to moving to 2D over the quasi-1D Lieb lattice, but we \mbox{include} one example here as a proof of concept.  A diagram of the \mbox{arrangement} of the 2D Lieb lattice we modelled is shown in Fig.~\ref{Lieb2Ddiag}.  The system considered is a $5\times5$ unit cell Lieb \mbox{lattice} (65 sites) with flat edges on all sides: the right most unit cells have no $A$ sites, while the bottom unit cells have no $C$ sites.  We focus on producing enhanced antibunching in the centremost $B$ site, in unit cell $(3,3)$ when labelling the unit cells by their location $(x,y)$.  We drive the system in a symmetrical way via the two closest $C$ sites, i.e.~$(3,3)C$ and $(3,4)C$, with a coherent drive of strength $F$.  The values of $n$ and $g^{(2)}(0)$ across the lattice for the parameters $U = 0.1\gamma$, $\Delta = -0.2\gamma$, $F = J = 3.0\gamma$ are shown in Fig.~\ref{Lieb2Dng2}.  With these parameters, the central $B$ site achieves $g^{(2)}(0) = 0.65$ with an occupation of $n = 0.23$.  Some other sites achieve stronger antibunching, but with correspondingly much smaller occupations.

\section{Conclusions}

We have used the positive-P method to explore various \mbox{situations} in which interesting and potentially exploitable quantum correlations can be induced by local driving in DDBH Lieb lattices through interference effects.  We have \mbox{focused} primarily on how antibunching ($g^{(2)}(0) < 1$) can be produced on one of the $B$ sites using a localised drive on the adjacent $C$ site.  In analogy to the \mbox{previously} discovered unconventional photon blockade, significant antibunching on the relevant $B$ site results when the parameters are tuned such that interference of the hopping terms from adjacent sites eliminates the amplitude for two bosons to occupy that site, while maintaining some possibility of single boson occupation.  

This can also be considered a counterpart to the mechanism of bunching in driven-dissipative Lieb lattices observed previously \cite{Casteels16}, where destructive interference of hopping in the lattice eliminates the single occupation amplitude on the dark sites, but leaves some amplitude for double occupations.  To support this view, in section \ref{section:FB} it was seen that destroying the flat band of the Lieb lattice also has the effect of destroying the enhanced antibunching.  This is not to claim that the enhanced antibunching in the locally driven Lieb lattice is a signature of flat band physics, since, as shown in \ref{section:3siteUPB}, it can occur similarly in the case of a single unit cell, where it no longer makes sense to consider the system to have the band structure of a full Lieb lattice.  Rather, the connection is that both the flat band and the enhanced antibunching share a common physical origin: destructive interference on the $B$ sites of the Lieb lattice.  The flat band is produced by destructive interference eliminating the occupation on, and therefore also hopping through, the $B$ sites, which in turn prevents bosons from moving through the lattice, giving a flat dispersion (no kinetic energy); on the other hand, the enhanced antibunching, as in the UPB, is produced by destructive interference eliminating the probability of multiple occupation of the relevant $B$ site.  The investigation of section \ref{section:FB} does also imply that this effect may be somewhat sensitive to variations in the \mbox{local} parameters of the lattice; however, in such scenarios it may also be possible to compensate by carefully tuning the drive or even driving additional sites, similar to \cite{PhysRevA.88.033836,PhysRevA.96.053810}, in order to re-establish the interference.  

We optimised the parameters of the model to maximise the ability to observe antibunching in polariton micropillar \mbox{lattices.}  Using the scalability of the positive-P method to \mbox{allow} for the simulation of extended lattices, we have shown there are some advantages to using larger structures for this \mbox{purpose} compared to the traditional 2 site UPB, or even our 3 site UPB of Section \ref{section:3siteUPB}.  Larger structures can both increase the period of oscillations in optimal conditions, as well as \mbox{reduce} the size of the initial peaks of $g^{(2)}(\tau)$ either side of the \mbox{minimum} at $\tau = 0$, as can be seen in the plots of oscillations in $g^{(2)}(\tau)$ throughout. Both effects aid in observing antibunching when using real detectors with finite time \mbox{resolution.}  Applying \mbox{additional} driving terms on other sites in the lattice can also be used to modify the interference with the aim of making the antibunching signal better suited for this purpose.  

Beyond this work, the results presented here are by no means exhaustive of all the possible localised driving schemes that one could investigate.  This work broadly opens the way to further exploration of different arrangements of drives and lattice geometries to find other situations in which antibunching or other interesting correlated states might be generated by \mbox{interference} effects in driven-dissipative quantum systems.  As an example, while preparing this manuscript we were made aware of a UPB-like effect recently demonstrated for a 4 site ring \cite{wang2025}. 


\acknowledgments{We thank the experimental groups of Alberto Amo, Jacqueline Bloch and Sylvain Ravets for helpful discussions.  A.F. acknowledges the support from the EUCENTRAL project. This project is funded by the European Union under Horizon Europe (project No. 101186579).  A.F. and M.H.S. acknowledge financial support from EPSRC (Grants No.~EP/S019669/1, No.~EP/V026496/1 and No.~EP/R04399X/1) and QuantERA InterPol.  P.D. acknowledges support by the National Science Centre (Poland) grant No.~2018/31/B/ST2/01871.  M.M. acknowledges support by the National Science Centre (Poland) grants No.~2017/25/Z/ST3/03032 (under the QuantERA program) and No.~2021/43/B/ST3/00752.}

\appendix
\section{Analytical optimisation of 3 site unconventional photon blockade}
\label{appendix:3siteUPB}

This appendix details the analytical solution of the 3 site DDBH model in the weak driving limit, used to calculate the parameters for optimal antibunching of the central site in section \ref{section:3siteUPB}.  The calculation proceeds as follows: first we begin by taking an ansatz for the state of the total system as a superposition of Fock states for up to two bosons distributed throughout the lattice
\begin{align}
|\psi\rangle =&\, C_{000}|000\rangle + e^{-i\omega_p t}\left(C_{100}|100\rangle + C_{010}|010\rangle \right. \nonumber \\
& \left. + C_{001}|001\rangle \right) +e^{-2i\omega_p t}\left(C_{200}|200\rangle \right. \nonumber \\
&\left. +C_{020}|020\rangle + C_{002}|002\rangle + C_{110}|110\rangle \right. \nonumber \\
&\left. + C_{101}|101\rangle + C_{011}|011\rangle \right) \, ,
\end{align}
with the amplitudes obeying a hierarchy 
\beqy
&C_{000} \gg C_{100},C_{010},C_{001} \gg \nonumber \\& C_{200},C_{020},C_{002},C_{110},C_{101},C_{011}\, , \label{Csizes}
\eeqy
due to the weak driving approximation.  This will be used to further approximate the density matrix $\hat{\rho}$ of the system, by keeping only up to linear order in the amplitudes other than $C_{000}$, leading to 
\beqy
\hat{\rho} = |\psi\rangle\langle\psi| \approx |C_{000}|^2|000\rangle\langle000| + \nonumber \\ C^*_{000}|\psi'\rangle\langle000| + C_{000}|000\rangle\langle\psi'|\, , \label{rhoApprox}
\eeqy
where $|\psi'\rangle = |\psi\rangle - C_{000}|000\rangle$.  

 We will take this approximation of $\hat{\rho}$, and substitute it into the master equation for the system.  For a general DDBH lattice and in the frame of the coherent drive, this is given by \eqref{ME_DDBH} with the Hamiltonian \eqref{hamBH}, but for this purpose we will rewrite it in the lab frame for this specific three site chain as
\begin{subequations}
\beqy
\dfrac{\partial \hat{\rho}}{\partial t} &=& -i[\hat{H}_{\rm 3site}, \hat{\rho}] \nonumber \\ &&+\dfrac{\gamma}{2}\sum_{j=1}^3 \left[ 2\hat{a}_j\hat{\rho}\hat{a}^\dagger_j -\hat{a}^\dagger_j\hat{a}_j\hat{\rho} - \hat{\rho}\hat{a}^\dagger_j\hat{a}_j \right] \, , \\
\nonumber \\
\hat{H}_{\rm 3site} &=& \sum_{j=1}^3 \left[ \omega\hat{a}^\dagger_j\hat{a}_j +\dfrac{U}{2}\hat{a}^\dagger_j\hat{a}^\dagger_j\hat{a}_j\hat{a}_j \right] \nonumber \\ &&+ Fe^{-i\omega_p t}\hat{a}^\dagger_1 + F^*e^{i\omega_p t}\hat{a}_j \nonumber \\
&&+J\left(\hat{a}^\dagger_1\hat{a}_2  + \hat{a}^\dagger_2\hat{a}_1 + \hat{a}^\dagger_2\hat{a}_3 + \hat{a}^\dagger_3\hat{a}_2 \right) \, .
\eeqy
\end{subequations}
Using \eqref{rhoApprox} and the properties of the creation and annihilation operators, this reduces to
\beqy
\lefteqn{\dfrac{\partial C^*_{000}}{\partial t} |\psi\rangle\langle000| + C^*_{000}\dfrac{\partial |\psi\rangle}{\partial t}\langle000| =} \nonumber \\&\qquad C^*_{000} \left( -i\hat{H}_{\rm 3site}|\psi\rangle\langle000| - \dfrac{\gamma}{2}\sum_{j=1}^3  \hat{a}^\dagger_j\hat{a}_j |\psi\rangle\langle000| \right) \nonumber \\
\lefteqn{\Rightarrow\, \dfrac{\partial C^*_{000}}{\partial t} |\psi\rangle + C^*_{000}\dfrac{\partial |\psi\rangle}{\partial t} =} &\nonumber \\& C^*_{000} \left( -i\hat{H}_{\rm 3site}|\psi\rangle - \dfrac{\gamma}{2}\sum_{j=1}^3  \hat{a}^\dagger_j\hat{a}_j |\psi\rangle \right) \, ,
\eeqy
and if we consider the steady state, i.e.~$\frac{\partial C_{abc}}{\partial t} = 0$ for all the amplitudes $C_{abc}$, then this further simplifies to
\beq
\frac{\partial |\psi\rangle}{\partial t} =  -i\hat{H}_{\rm 3site}|\psi\rangle - \frac{\gamma}{2}\sum_{j=1}^3  \hat{a}^\dagger_j\hat{a}_j |\psi\rangle  \, . \label{reducedSE}
\eeq
Furthermore, in the steady state, the left hand side also can be expanded as 
\beqy
\dfrac{\partial |\psi\rangle}{\partial t} &=& -i\omega_pe^{-i\omega_p t}\left(C_{100}|100\rangle + C_{010}|010\rangle + C_{001}|001\rangle \right) \nonumber \\&&
 -2i\omega_pe^{-2i\omega_p t}\left(C_{200}|200\rangle + C_{020}|020\rangle + C_{002}|002\rangle \right. \nonumber \\&&
\left. + \,C_{110}|110\rangle + C_{101}|101\rangle + C_{011}|011\rangle \right) \, .
\eeqy
Substituting this into \eqref{reducedSE} and comparing the coefficients of each Fock state $|abc\rangle$, leads to a set of nine simultaneous equations for the amplitudes $C_{abc}$.  These are 
\begin{subequations}
\begin{eqnarray}
-i\omega_pC_{100} &=& -i\omega C_{100} -iJC_{010}  -iFC_{000} -\dfrac{\gamma}{2}C_{100} \nonumber \\
\Rightarrow\, 0 &=& \left( -\Delta - \dfrac{i\gamma}{2} \right)C_{100} + JC_{010} + FC_{000}, \qquad\label{coeffeq1} 
\end{eqnarray}
\begin{eqnarray}
-i\omega_pC_{010} &=& -i\omega C_{010} -iJC_{100}  -iJC_{001} -\dfrac{\gamma}{2}C_{010} \nonumber \\
\Rightarrow\, 0 &=& \left( -\Delta - \dfrac{i\gamma}{2} \right)C_{010} + J\left(C_{100} + C_{001}\right), \qquad\label{coeffeq2} 
\end{eqnarray}
\begin{eqnarray}
-i\omega_pC_{001} &=& -i\omega C_{001} -iJC_{010} -\frac{\gamma}{2}C_{001} \nonumber \\
\Rightarrow\, 0 &=& \left( -\Delta - \frac{i\gamma}{2} \right)C_{001} + JC_{010}, \qquad\label{coeffeq3} 
\end{eqnarray}
\begin{eqnarray}
-2i\omega_pC_{200} &=& -2i\omega C_{200} -i\sqrt{2}JC_{110} \nonumber\\&&-i\sqrt{2}FC_{100} -\gamma\,C_{200} - iUC_{200} \nonumber \\
\Rightarrow\, 0 &=& 2\left( -\Delta + \frac{U}{2} - \frac{i\gamma}{2} \right)C_{200} \nonumber\\
&&+ \sqrt{2}JC_{110} + \sqrt{2}FC_{100},\qquad\label{coeffeq4} 
\end{eqnarray}
\begin{eqnarray}
-2i\omega_pC_{020} &=& -2i\omega C_{020} -i\sqrt{2}JC_{110} \nonumber\\&&-i\sqrt{2}JC_{011}  -\gamma\,C_{020} - iUC_{020} \nonumber \\
\Rightarrow\, 0 &=& 2\left( -\Delta + \frac{U}{2} - \frac{i\gamma}{2} \right)C_{020}  \nonumber\\&&+ \sqrt{2}J\left(C_{110} + C_{011}\right),\qquad\label{coeffeq5} 
\end{eqnarray}
\begin{eqnarray}
&-2i\omega_pC_{002} = -2i\omega C_{002} -i\sqrt{2}JC_{011} -\gamma\,C_{002} - iUC_{002} \nonumber \\
&\Rightarrow\, 0 = 2\left( -\Delta + \frac{U}{2} - \frac{i\gamma}{2} \right)C_{002} + \sqrt{2}JC_{011},\qquad \label{coeffeq6} 
\end{eqnarray}
\begin{eqnarray}
-2i\omega_pC_{110} &=& -2i\omega C_{110} -i\sqrt{2}JC_{200} \nonumber\\&&
-i\sqrt{2}JC_{020} -iJC_{101} -iFC_{010} -\gamma\,C_{110} \nonumber \\
\Rightarrow\, 0 &=& 2\left( -\Delta - \frac{i\gamma}{2} \right)C_{110}  + FC_{010} \nonumber\\&&
+ J\left(C_{101} +\sqrt{2}\left[C_{200} + C_{020}\right]\right),\qquad\label{coeffeq7} 
\end{eqnarray}
\begin{eqnarray}
-2i\omega_pC_{101} &=& -2i\omega C_{101} -iJC_{110}  -iJC_{011} \nonumber\\
&&-iFC_{001} -\gamma\,C_{101} \nonumber \\
\Rightarrow\, 0 &=& 2\left( -\Delta - \frac{i\gamma}{2} \right)C_{101}  \nonumber\\&&+ J\left(C_{110} + C_{011}\right) + FC_{001},\qquad\label{coeffeq8} 
\end{eqnarray}
\begin{eqnarray}
-2i\omega_pC_{011} &=& -2i\omega C_{011} -i\sqrt{2}JC_{002}  -i\sqrt{2}JC_{020} \nonumber\\
&&-iJC_{101} -\gamma\,C_{011} \nonumber \\
\Rightarrow\, 0 &=& 2\left( -\Delta - \dfrac{i\gamma}{2} \right)C_{011} \nonumber\\&&
+ J\left(C_{101} +\sqrt{2}\left[C_{002} + C_{020}\right]\right),\qquad \label{coeffeq9}
\end{eqnarray}
\end{subequations}
where for the single occupation amplitudes \eqref{coeffeq1}-\eqref{coeffeq3}, we have used the approximation \eqref{Csizes} to neglect any dependence of these on the two boson amplitudes, and $\Delta = \omega_p - \omega$.  Rearranging \eqref{coeffeq3}, gives 
\beq
C_{001} = \frac{JC_{010}}{\Delta +\frac{i\gamma}{2}} \, , \label{coeffeq3r}
\eeq
which can then be substituted into \eqref{coeffeq2} to give
\begin{align}
0 =& \left( -\Delta - \frac{i\gamma}{2} \right)C_{010} + JC_{100} + \frac{J^2 C_{010}}{\Delta +\frac{i\gamma}{2}} \nonumber \\
\Rightarrow\, C_{100} =& \left( \frac{\Delta +\frac{i\gamma}{2}}{J} - \frac{J}{\Delta +\frac{i\gamma}{2}} \right) C_{010} \, .  \label{coeffeq2r}
\end{align}
Similarly for \eqref{coeffeq6}, 
\beq
C_{002} = \frac{JC_{011}}{\left(\Delta - \frac{U}{2} +\frac{i\gamma}{2}\right)\sqrt{2}} \, , \label{coeffeq6r}
\eeq
and substituting this into  \eqref{coeffeq9} gives
\begin{align}
0 =& 2\left( -\Delta - \frac{i\gamma}{2} \right)C_{011} + JC_{101} \nonumber \\&+\sqrt{2}JC_{020} + \frac{J^2 C_{011}}{\Delta -\frac{U}{2} +\frac{i\gamma}{2}} \nonumber \\
\Rightarrow\,& C_{101} +\sqrt{2}C_{020} = \nonumber \\& \left( \frac{2\left(\Delta +\frac{i\gamma}{2}\right)}{J} - \frac{J}{\Delta - \frac{U}{2} +\frac{i\gamma}{2}}   \right) C_{011} \, .  \label{coeffeq9r}
\end{align}

To obtain the parameters for optimal antibunching on the central site, \mbox{$g^{(2)}_2(0) = 0$}, we can proceed by setting the amplitude for double occupation of that site \mbox{$C_{020} = 0$}.  As well as eliminating $C_{020}$ from \eqref{coeffeq9r} above, this also reduces \eqref{coeffeq5} to a simple relation $C_{110} = -C_{011}$, which when substituted into \eqref{coeffeq8} gives
\beq
C_{001} = \frac{2\left(\Delta +\frac{i\gamma}{2}\right)}{F}C_{101}\, .
\eeq
Applying \eqref{coeffeq3r} and \eqref{coeffeq2r} sequentially to this gives equations for the other two single boson amplitudes in terms of $C_{101}$
\begin{align}
C_{010} =&\, \frac{2\left(\Delta +\frac{i\gamma}{2}\right)^2}{JF}C_{101}\, , 
\end{align}
\begin{align}
C_{100} =&\, 2\left( \frac{\left(\Delta +\frac{i\gamma}{2}\right)^3}{J^2 F} - \frac{\Delta +\frac{i\gamma}{2}}{F} \right) C_{101}\, ,
\end{align}
the second of which, along with \eqref{coeffeq9r} to replace $C_{110} = -C_{011}$, can be substituted into  \eqref{coeffeq4} to give $C_{200}$ in terms of $C_{101}$, 
\beqy
&C_{200} = \dfrac{1}{\sqrt{2}}\left[ \dfrac{2\left(\Delta +\frac{i\gamma}{2}\right)^3}{J^2 \left( \Delta - \frac{U}{2} +\frac{i\gamma}{2}\right) } - \dfrac{2\left(\Delta +\frac{i\gamma}{2}\right)}{\Delta - \frac{U}{2} +\frac{i\gamma}{2}} \right. \nonumber \\& \left.
- \dfrac{J^2}{2\left(\Delta +\frac{i\gamma}{2}\right)\left( \Delta - \frac{U}{2} +\frac{i\gamma}{2}\right) - J^2} \right] C_{101} \, .
\eeqy
Finally, we use \eqref{coeffeq7} and, after setting \mbox{$C_{020} = 0$}, rewrite all other amplitudes in terms of $C_{101}$ by substituting in the respective equations above.  This leads to an equation with just the parameters of the model and $C_{101}$, 
\begin{align}
&\left[ \frac{2J\left(\Delta +\frac{i\gamma}{2}\right)\left( \Delta - \frac{U}{2} +\frac{i\gamma}{2}\right)}{2\left(\Delta +\frac{i\gamma}{2}\right)\left( \Delta - \frac{U}{2} +\frac{i\gamma}{2}\right) - J^2} + J + \frac{2\left(\Delta +\frac{i\gamma}{2}\right)^3}{J \left( \Delta - \frac{U}{2} +\frac{i\gamma}{2}\right) } \right. \nonumber \\ &\quad\left. - \frac{2J\left(\Delta +\frac{i\gamma}{2}\right)}{\Delta - \frac{U}{2} +\frac{i\gamma}{2}} - \frac{J^3}{2\left(\Delta +\frac{i\gamma}{2}\right)\left( \Delta - \frac{U}{2} +\frac{i\gamma}{2}\right) - J^2} \right. \nonumber \\&
\quad\left. + \frac{2\left(\Delta +\frac{i\gamma}{2}\right)^2}{J} \right] C_{101} = 0\, ,
\end{align}
which can be further simplified as we need $C_{101} \neq 0$, otherwise all the amplitudes will become 0, allowing $C_{101}$ to be divided through to leave an equation relating only the physical parameters of the DDBH model.  After some rearranging of terms, this takes the form
\beq
4\left(\Delta +\frac{i\gamma}{2}\right)^3 -U\left(\Delta +\frac{i\gamma}{2}\right)^2 -J^2U = 0\, .
\eeq
Expanding the powers of $\left(\Delta +\frac{i\gamma}{2}\right)$, we can split this into two equations for the real and imaginary parts, as given by equations \eqref{opteq}.

\begin{figure}[h]
\begin{center}
\includegraphics[width = \columnwidth]{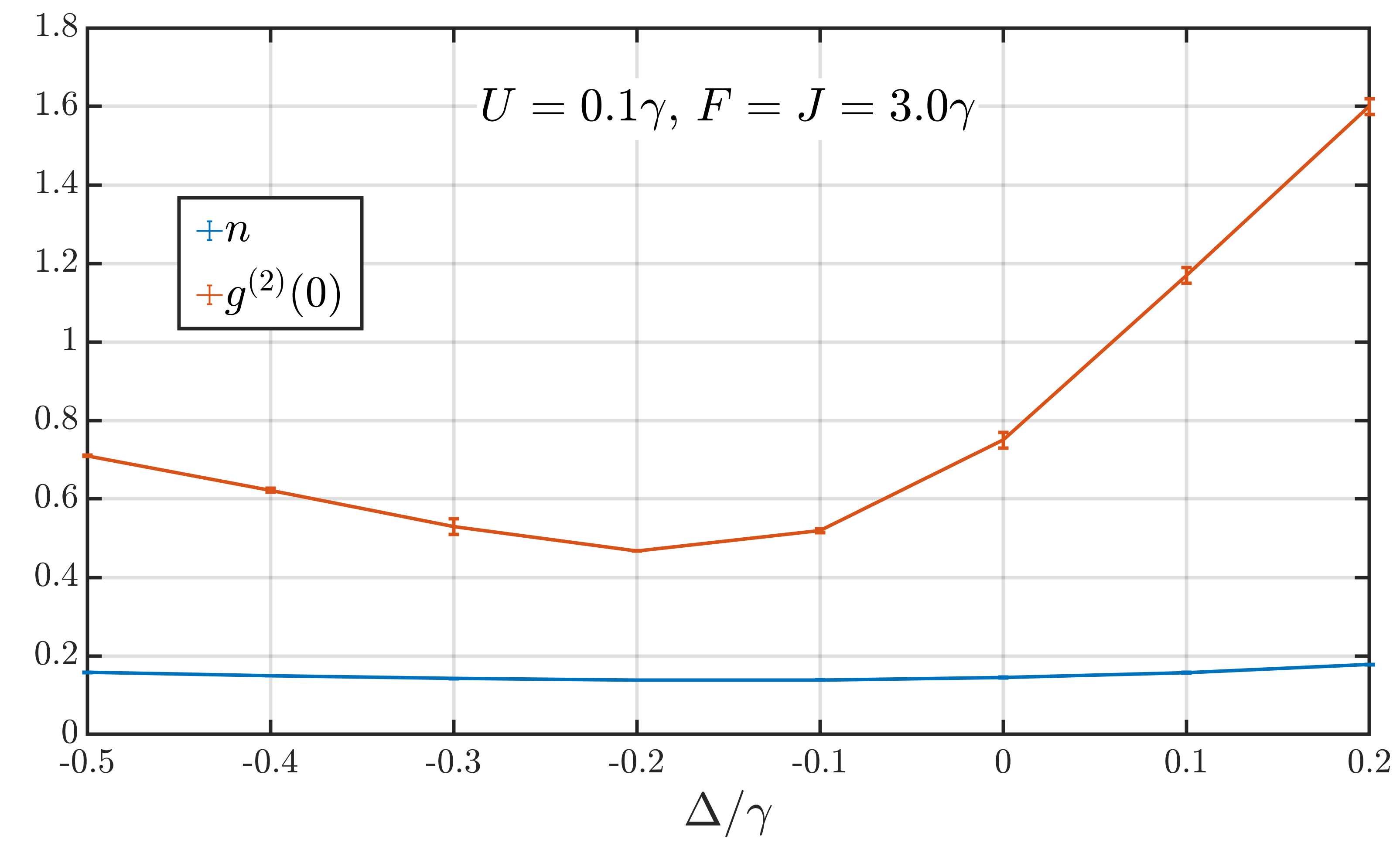}
\caption{Variation in $n$ and $g^{(2)}(0)$ of site $3B$ with $\Delta$.  Other parameters are fixed as $U = 0.1\gamma$, $F = J = 3.0\gamma$.  \label{LiebLpDvar}}
\end{center}
\end{figure}

\begin{figure}[h]
\begin{center}
\includegraphics[width = \columnwidth]{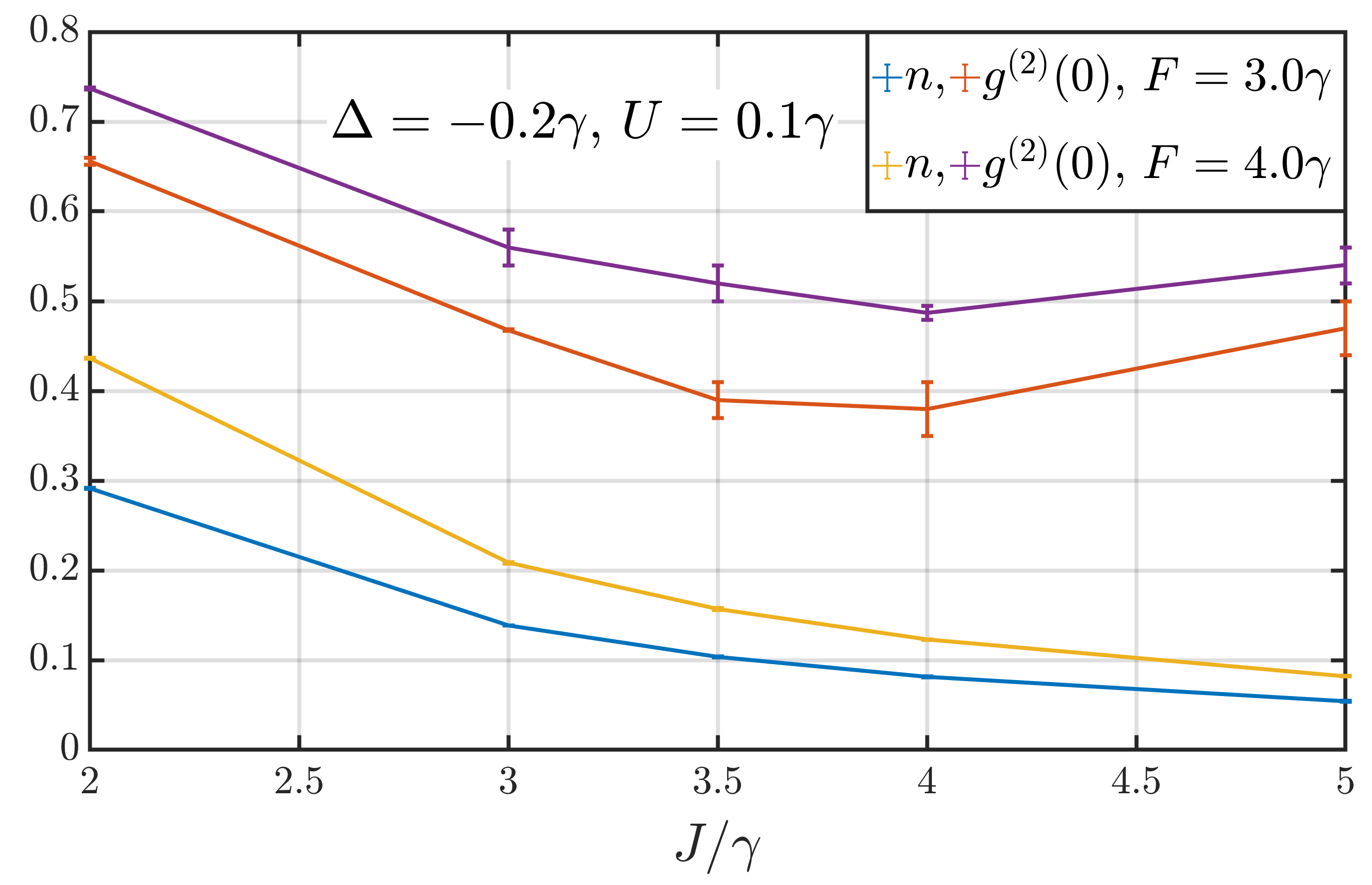}
\caption{Variation in $n$ and $g^{(2)}(0)$ of site $3B$ with $J$ at fixed $F =3.0\gamma$ and $F = 4.0\gamma$.  Other parameters are fixed as $U = 0.1\gamma$, $\Delta = -0.2\gamma$.  \label{LiebLpJvar}}
\end{center}
\end{figure}

\section{Optimising parameters for observability in quasi-1D Lieb lattices}\label{appendix:LiebOpt}

\begin{figure}[t]
\includegraphics[width = \columnwidth]{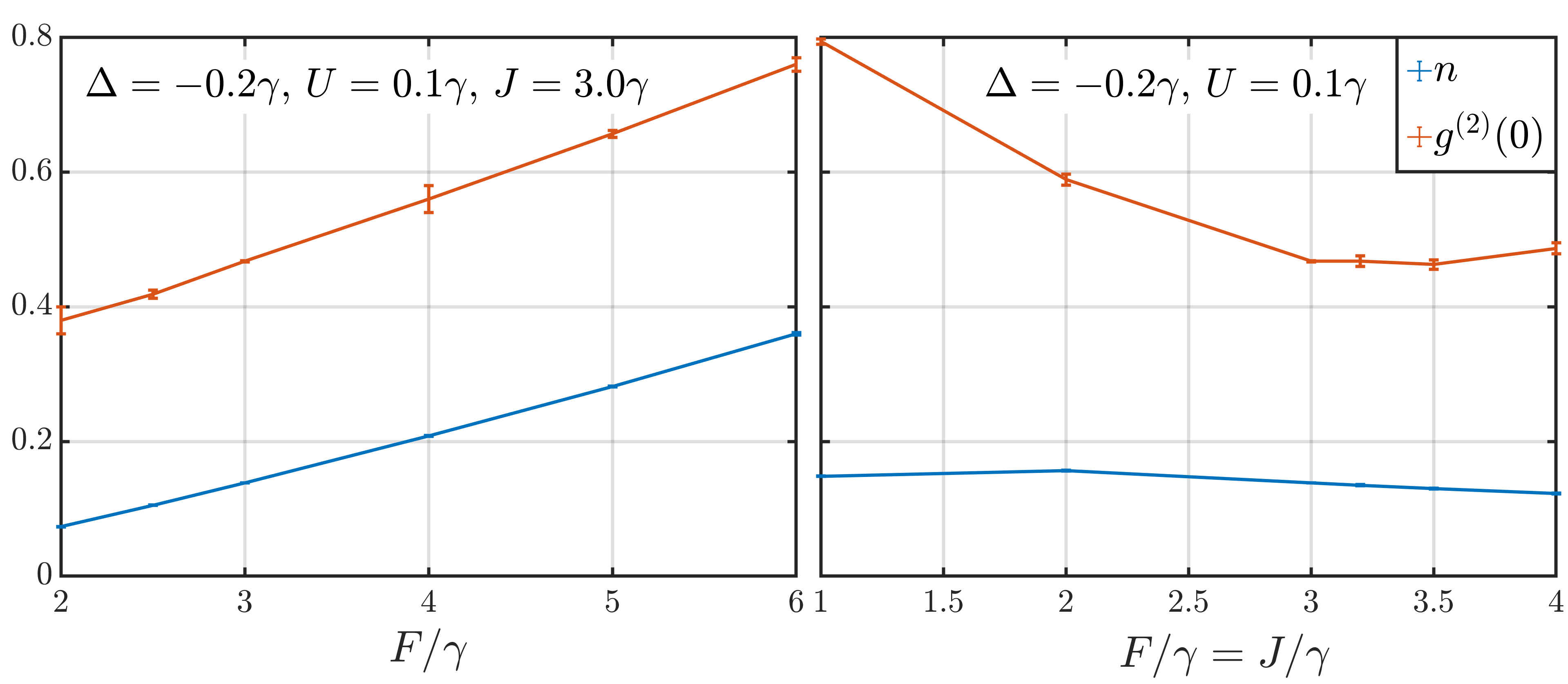}
\caption{Variation in $n$ and $g^{(2)}(0)$ of site $3B$ with drive strength $F$ at fixed $J =3.0\gamma$ \mbox{(left panel)} and when maintaining relation $F = J$ (right panel).  Other parameters are fixed as $U = 0.1\gamma$, $\Delta = -0.2\gamma$.  \label{LiebLpFvar}}
\end{figure}

\begin{figure}[b]
\centering
\includegraphics[width = \columnwidth]{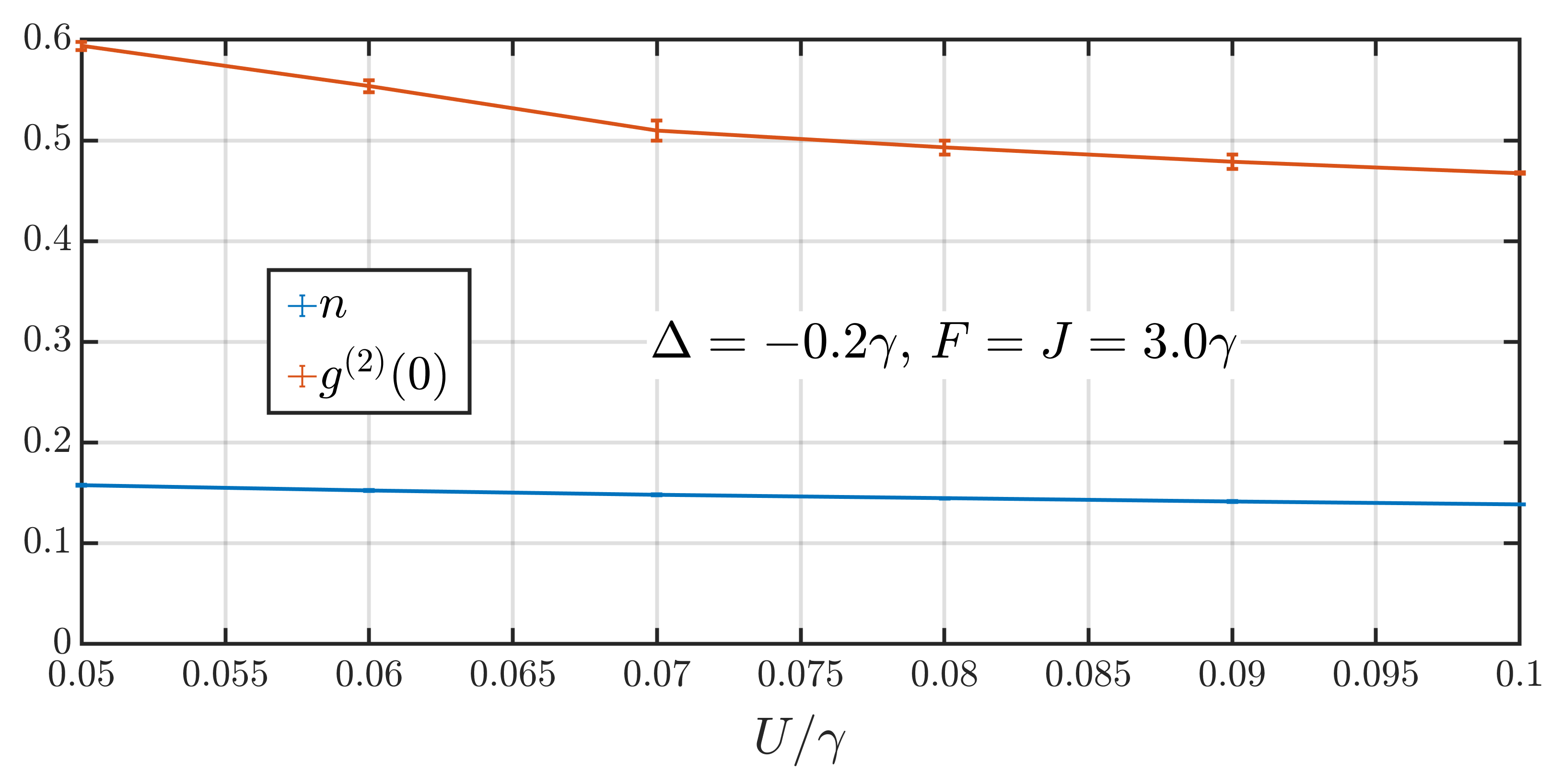}
\caption{Variation in $n$ and $g^{(2)}(0)$ of site $3B$ with $U$.  Other parameters are fixed as $\Delta = -0.2\gamma$,  $F = J = 3.0\gamma$.  \label{LiebLpUvar}}
\end{figure}

Here, we investigate how the antibunching on the target site $3B$ and other factors that would affect its experimental observability, are affected by varying the physical parameters of the DDBH model, $\Delta$, $F$, $J$, and $U$ for the system of Figs.~\ref{LiebLpDiag} and~\ref{LiebBgDiag} in sections \ref{section:LiebLp}-\ref{section:LiebBG}; $\gamma$ is taken as the energy scale.  We optimise this system by the following priorities: firstly, the aim is to minimise the value of $g^{(2)}(0)$ on site $3B$, however, the occupation of site $3B$ must also not be so small as to make it unrealistic to actually perform $g^{(2)}$ measurements of that site; we take $n > 0.1$ as a lower bound.  A third possible consideration is to make the period of oscillations in $g^{(2)}(\tau)$ as large as possible, which would be achieved by making $J$ as small as possible, without too heavily sacrificing the degree of antibunching achieved.  To begin with, we focus on the optimisation of parameters in this way for the single drive setup of section \ref{section:LiebLp}.  We will show how the parameters used in Fig.~\ref{LiebLpng2all} are chosen as a good compromise between all these factors.  

The first consideration is the dependence on the local energy detuning $\Delta$, which is given in Fig.~\ref{LiebLpDvar}.  It can be seen that for this setup, $n$ does not depend strongly on $\Delta$.  On the other hand, a clear minimum for $g^{(2)}(0)$ is seen at around $\Delta = -0.2\gamma$.  Furthermore, positive values of $\Delta$ destroy the desired effect completely by eventually switching $g^{(2)}(0)$ from antibunched $g^{(2)}(0) < 1$ to bunched $g^{(2)}(0) > 1$.  This is consistent with our observation in section \ref{section:3siteUPB} that only negative solutions for $\Delta_{opt}$ gave real solutions for $J_{opt}$.  

Fixing to $\Delta = -0.2\gamma$, the variation with $J$ is shown in Fig.~\ref{LiebLpJvar} for both \mbox{$F = 3.0\gamma$} and \mbox{$F = 4.0\gamma$}.  $g^{(2)}(0)$ reaches a minimum at around $J = 4.0\gamma$ in both cases, although $n$ also decreases with increasing $J$.  Using a slightly smaller value such as $J = 3.0\gamma$ can have practical benefits of both increasing $n$ on the relevant site as well as increasing the period of oscillations in $g^{(2)}(\tau)$.  

\begin{figure}[t]
\begin{center}
\includegraphics[width = \columnwidth]{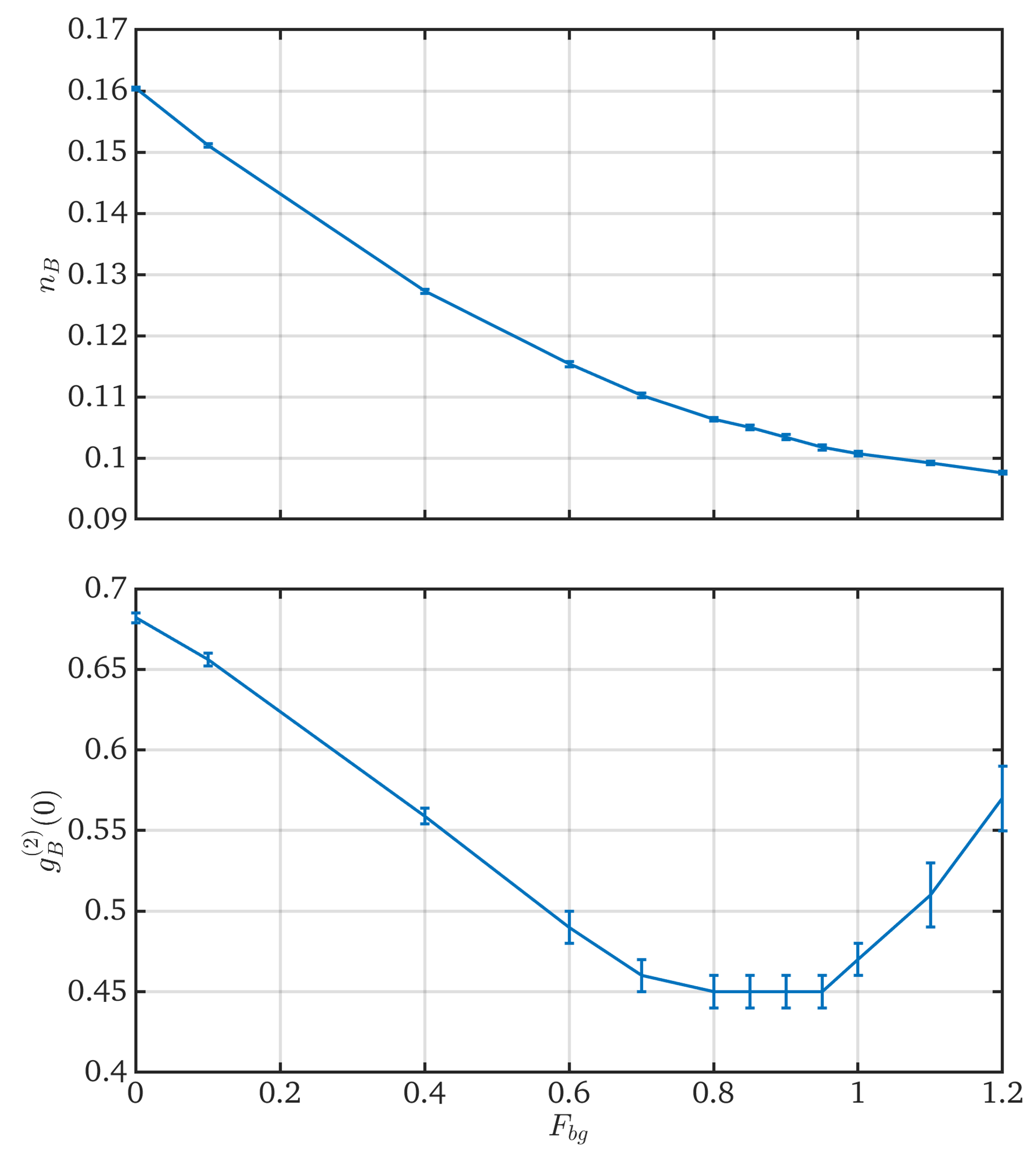}
\caption{Variation in $n$ (top) and $g^{(2)}(0)$ (bottom) of site $3B$ with the background drive strength $F_{bg}$.  Other parameters are fixed as $\Delta = -0.2\gamma$, $U = 0.1\gamma$, $F = J = 1.5\gamma$.  \label{FbgOpt}}
\end{center}
\end{figure}

Similar to as was demonstrated for the single unit cell in Fig.~\ref{3sO_PPg2t}, increasing the coherent drive $F$ while fixing the other parameters has the effect of linearly increasing both $n$ and $g^{(2)}(0)$ of the relevant site, as shown in the left panel of Fig.~\ref{LiebLpFvar}.  As such, $F$ should be made as small as possible while maintaining sufficiently large $n$.  Perhaps a more useful comparison is then given by the right panel of Fig.~\ref{LiebLpFvar}, where instead of only varying $F$, $F$ and $J$ are varied together keeping $F = J$ and all other parameters constant.  This choice effects the value of $g^{(2)}(0)$ without significantly altering $n$.  Here, a minimum value of $g^{(2)}(0)$ is achieved between $F = J = 3.0\gamma$ and $F = J = 3.5\gamma$, again with the smallest value of $J$ within this range being ideal due to maximising the period of $g^{(2)}(\tau)$.  In Fig.~\ref{LiebLpUvar}, we show that neither $n$ nor $g^{(2)}(0)$ on site $3B$ are strongly affected by slight variations in the interaction strength $U$ in this setup.  This is helpful for the possibility of reproducing these results in polariton micropillar experiments, as it means there is no need to exactly match the value of $U = 0.1\gamma$ that we use here.  

Finally, we investigate optimising with the strength of the additional background drive included in the setup of section \ref{section:LiebBG}.  We focus on parameter values $\Delta = -0.2\gamma$, $U = 0.1\gamma$, $F = J = 1.5\gamma$, for which the variation in $n$ and $g^{(2)}(0)$ of site $3B$ with the background drive strength $F_{bg}$ is shown in \mbox{Fig.~\ref{FbgOpt}.}  It can be seen that a minimum value of $g^{(2)}(0) = 0.45$, comparable to that achieved in the ideal case chosen in section \ref{section:LiebLp}, is reached for values of $F_{bg}$ between $0.8\gamma$ and $0.95\gamma$.  While increasing $F_{bg}$ does also have the effect of slowly decreasing the occupation $n$ of $3B$, the occupation for $F_{bg} = 0.8\gamma$ is still similar to that for the example of section \ref{section:LiebLp}.  As such, the addition of the background drive allows equivalent results for $n$ and $g^{(2)}(0)$ to be achieved at half the value of $J$.

\bibliography{Bibliographies/Books,Bibliographies/Miscellarticles,Bibliographies/PhysRevLett,Bibliographies/PhysRevB,Bibliographies/ArXiv,Bibliographies/OtherPhysRev,Bibliographies/Nature_all,Bibliographies/artnew}

\end{document}